\documentclass[aip,jmp,amsmath,amssymb,reprint,onecolumn]{revtex4-1}

\usepackage{graphicx}
\usepackage{bm}

\usepackage[utf8]{inputenc}
\usepackage[T1]{fontenc}
\usepackage{mathptmx}
\usepackage{etoolbox}

\usepackage{geometry}
\usepackage{amsthm}
\usepackage[colorlinks=true]{hyperref}

\newtheorem{thm}{Theorem}

\newtheorem*{thm*}{Theorem}
\newtheorem{cor}{Corollary}

\newtheorem{lem}{Lemma}

\newtheorem{prop}{Proposition}

\theoremstyle{definition}
\newtheorem{defn}{Definition}

\theoremstyle{remark}
\newtheorem{rem}{Remark}

\makeatletter
\def\@email#1#2{%
 \endgroup
 \patchcmd{\titleblock@produce}
  {\frontmatter@RRAPformat}
  {\frontmatter@RRAPformat{\produce@RRAP{*#1\href{mailto:#2}{#2}}}\frontmatter@RRAPformat}
  {}{}
}%
\makeatother
\begin{document}

\title{The generalized strong subadditivity of the von Neumann entropy for bosonic quantum systems}
\author{Giacomo De Palma}
 \affiliation{University of Bologna, Bologna, Italy}
 \affiliation{Scuola Normale Superiore, Pisa, Italy}
 \email{giacomo.depalma@unibo.it}
\author{Dario Trevisan}%
 \email{dario.trevisan@unipi.it}
\affiliation{University of Pisa, Pisa, Italy}

\date{\today}

\begin{abstract}
We prove a generalization of the strong subadditivity of the von Neumann entropy for bosonic quantum Gaussian systems.
Such generalization determines the minimum values of linear combinations of the entropies of subsystems associated to arbitrary linear functions of the quadratures, and holds for arbitrary quantum states including the scenario where the entropies are conditioned on a memory quantum system.
We apply our result to prove new entropic uncertainty relations with quantum memory, a generalization of the quantum Entropy Power Inequality, and the linear time scaling of the entanglement entropy produced by quadratic Hamiltonians.
\end{abstract}

\maketitle

\section{Introduction and main result}
Let $X$ be a random variable with values in $\mathbb{R}^n$ and probability density $p$ with normalization\footnote{We have included the factor $\left(2\pi\right)^{-\frac{n}{2}}$ in the integration measure to avoid having it in the probability densities of Gaussian random variables.}
\begin{equation}
\int_{\mathbb{R}^n}p(x)\,\frac{\mathrm{d}^nx}{\left(2\pi\right)^\frac{n}{2}} = 1\,.
\end{equation}
The \emph{Shannon differential entropy} of $X$ is [\onlinecite{cover2012elements}]
\begin{equation}
S(X) = - \mathbb{E}_X\ln p(X)\,,
\end{equation}
and quantifies the uncertainty of $X$, \emph{i.e.}, the information that is obtained when the value of $X$ is revealed.

Given the entropy of a random variable $X$, what can we say about the entropies of the linear functions of $X$?
The question can be formalized as follows:
\begin{defn}[Brascamp--Lieb datum]
A \emph{Brascamp--Lieb datum} with dimension $(n,\mathbf{n})\in\mathbb{N}\times\mathbb{N}^K$ is an ordered set $\mathbf{B}=(B_1,\,\ldots,\,B_K)$, where each $B_i$ is an $n_i\times n$ real matrix with rank $n_i$.
For any $\mathbf{p}\in\mathbb{R}_{\ge0}^K$, the \emph{Brascamp--Lieb constant} of the pair $(\mathbf{B},\mathbf{p})$ is
\begin{equation}\label{eq:fX}
f(\mathbf{B},\mathbf{p}) = \sup_X \left(S(X) - \sum_{i=1}^K p_i\,S(B_iX)\right)\,,
\end{equation}
where the supremum is taken over all the random variables $X$ with values in $\mathbb{R}^n$ and finite Shannon differential entropy.
\end{defn}
The definition of $f(\mathbf{B},\mathbf{p})$ can be rephrased as the following inequality, which generalizes the subadditivity of the Shannon differential entropy to the entropies of the linear functions of $X$:
\begin{equation}\label{eq:GSSA}
S(X) \le \sum_{i=1}^K p_i\,S(B_iX) + f(\mathbf{B},\mathbf{p})\,.
\end{equation}
The inequality \eqref{eq:GSSA} is dual to a functional inequality called Brascamp--Lieb inequality (see \autoref{app:BL}), hence the name Brascamp--Lieb datum.
A seminal result by Lieb [\onlinecite{lieb1990gaussian}] implies that the supremum in \eqref{eq:fX} can be restricted to centered Gaussian probability distributions, thus reducing the computation of $f(\mathbf{B},\mathbf{p})$ to the following optimization over covariance matrices (\autoref{thm:Lieb}):
\begin{equation}\label{eq:fdet}
f(\mathbf{B},\mathbf{p}) = \frac{1}{2}\sup_{\alpha\in\mathbb{R}^{n\times n}_{>0}}\left(\ln\det\alpha - \sum_{i=1}^Kp_i\ln\det\left(B_i\,\alpha\,B_i^T\right)\right)\,,
\end{equation}
where $\mathbb{R}^{n\times n}_{>0}$ denotes the set of the positive-definite $n\times n$ real matrices.

In this paper, we generalize the inequality \eqref{eq:GSSA} to bosonic quantum Gaussian systems.
A \emph{bosonic quantum Gaussian system} $X$ with $m$ modes provides a noncommutative generalization of $\mathbb{R}^{2m}$, where the $2m$ real coordinates are replaced by $2m$ linear operators $R_1^X,\,\ldots,\,R_{2m}^X$, called \emph{quadratures}, which act on the Hilbert space $L^2(\mathbb{R}^m)$ and satisfy the canonical commutation relations [\onlinecite{ferraro2005gaussian,weedbrook2012gaussian,serafini2017quantum,holevo2019quantum,hackl2020bosonic}]
\begin{equation}\label{eq:CCR}
\left[R_i^X,\,R_j^X\right] = \mathrm{i}\left(\Delta_{2m}\right)_{ij}\mathbb{I}\,,\qquad i,\,j=1,\,\ldots,\,2m\,,\qquad
\Delta_{2m} = \bigoplus_{i=1}^m \left(
                                  \begin{array}{cc}
                                    0 & 1 \\
                                    -1 & 0 \\
                                  \end{array}
                                \right)\,.
\end{equation}
For any $i=1,\,\ldots,\,m$, the $i$-th mode of $X$ is associated to the quadratures $R_{2i-1}^X$ and $R_{2i}^X$, which satisfy $\left[R_{2i-1}^X,\,R_{2i}^X\right] = \mathrm{i}\,\mathbb{I}$, and the quadratures associated to different modes commute.
The quantum counterpart of the probability distributions on $\mathbb{R}^{2m}$ are the \emph{quantum states} of $X$, \emph{i.e.}, the positive semidefinite linear operators with unit trace acting on $L^2(\mathbb{R}^m)$.
Bosonic quantum Gaussian systems provide the mathematical model for the electromagnetic radiation in the quantum regime and play a key role in quantum information theory, since most protocols for quantum communication are based on light traveling through optical fibers [\onlinecite{weedbrook2012gaussian}].

We will employ a conditioning quantum system $M$ associated to a separable Hilbert space $\mathcal{H}_M$.
A joint quantum state $\rho_{XM}$ of $XM$ is a quantum state acting on $L^2(\mathbb{R}^m)\otimes\mathcal{H}_M$.
We denote with $\rho_X$ and $\rho_M$ the marginal states of $\rho_{XM}$ on the subsystems $X$ and $M$, respectively.

We also need the notion of a joint state of a random variable and a quantum system.
If $X$ is a random variable with values in $\mathbb{R}^n$, a joint state of $XM$ is a collection of quantum states of $M$ $\left\{\rho_M(x):x\in\mathbb{R}^n\right\}$ such that the function $x\mapsto\rho_M(x)$ is measurable, the quantum system $M$ conditioned on $X=x$ has state $\rho_M(x)$ for any $x\in\mathbb{R}^n$, and the resulting marginal state of $M$ is $\rho_M = \mathbb{E}_X\,\rho_M(X)$.

Let $B$ be an $n\times2m$ real matrix with rank $n$ (with $n\le2m$).
To define the quantum counterpart of the linear transformation $Y = BX$, we define the $n$ quadratures
\begin{equation}\label{eq:ry}
R_i^Y = \sum_{j=1}^{2m} B_{ij}\,R_j^X\,,\qquad i=1,\,\ldots,\,n\,.
\end{equation}
We consider the following two cases:
\begin{description}
\item[Quantum case] If $n$ is even and $B$ satisfies
\begin{equation}\label{eq:DeltaY}
B\,\Delta_{2m}\,B^T = \Delta_n\,,
\end{equation}
the quadratures $R_1^Y,\,\ldots,\,R_n^Y$ satisfy the canonical commutation relations \eqref{eq:CCR} and are therefore unitarily equivalent to the quadratures $R_1^X,\,\ldots,\,R_n^X$, \emph{i.e.}, there exists a Gaussian symplectic unitary operator $U$ acting on $L^2(\mathbb{R}^m)$ such that $R_i^Y = U^\dag\,R_i^X\,U$ for any $i=1,\,\ldots,\,n$ (see \emph{e.g.} [\onlinecite[Section 5.1.2]{serafini2017quantum}]).
We denote with $\Phi_B$ the quantum channel acting on the quantum states of $X$ that applies the unitary operator $U$ and discards the last $m-n/2$ modes, therefore keeping the subsystem with $n/2$ modes associated to the quadratures $R_1^Y,\,\ldots,\,R_n^Y$.
We denote such subsystem with $Y$.
\item[Classical case] If $B$ satisfies
\begin{equation}\label{eq:Bcomm}
B\,\Delta_{2m}\,B^T = 0\,,
\end{equation}
the quadratures $R_1^Y,\,\ldots,\,R_n^Y$ commute and can be jointly measured.
We denote with $Y$ the random variable associated to the outcome of such measurement, and with $\Phi_B$ the quantum-to-classical channel acting on the quantum states of $X$ that outputs the probability distribution of $Y$.
In particular, \eqref{eq:Bcomm} is trivially satisfied if $B$ has rank one, \emph{i.e.}, if $n=1$, in which case $Y$ is the outcome of the measurement of a single quadrature.
\end{description}

The \emph{von Neumann entropy} of the quantum system $X$ in the state $\rho_X$ is [\onlinecite{nielsen2010quantum,wilde2017quantum,holevo2019quantum}]
\begin{equation}
S(X)(\rho_X) = -\mathrm{Tr}\left[\rho_X\ln\rho_X\right]\,,
\end{equation}
and quantifies the amount of noise or information contained in $X$.
When there is no ambiguity on the state $\rho_X$, we denote such entropy with $S(X)$.
The \emph{conditional von Neumann entropy} of $X$ given $M$ is [\onlinecite{wilde2017quantum,holevo2019quantum}]
\begin{equation}
S(X|M) = S(XM) - S(M)\,,
\end{equation}
provided that both $S(XM)$ and $S(M)$ are finite.
This condition always holds for the joint quantum states $\rho_{XM}$ of $XM$ such that $S(M)$ is finite and $\rho_X$ has finite energy, \emph{i.e.}, $\mathrm{Tr}\left[R_i^X\,\rho_X\,R_i^X\right]$ is finite for any $i=1,\,\ldots,\,2m$.
We denote the set of such states with $\mathcal{S}(XM)$.

In the case where $X$ is a random variable with values in $\mathbb{R}^n$, we define the \emph{conditional Shannon differential entropy} of $X$ given $M$ as
\begin{equation}
S(X|M) = S(M|X) + S(X) - S(M)\,,
\end{equation}
where
\begin{equation}
S(M|X)(\rho_{XM}) = \mathbb{E}_X \,S(M)(\rho_M(X))\,,
\end{equation}
provided that both $S(X)$ and $S(M)$ are finite.

\begin{defn}[Quantum Brascamp--Lieb datum]\label{defn:QBL}
A \emph{quantum Brascamp--Lieb datum} with dimension $(2m,\mathbf{n})\in2\mathbb{N}\times\mathbb{N}^K$ is a Brascamp--Lieb datum $\mathbf{B}$ where each $B_i$ satisfies either \eqref{eq:DeltaY} or \eqref{eq:Bcomm}.
For each $i=1,\,\ldots,\,K$, we denote with $Y_i$ the output system of $\Phi_{B_i}$, which is either a quantum Gaussian system with $n_i/2$ modes or a random variable with values in $\mathbb{R}^{n_i}$, depending on whether $B_i$ satisfies \eqref{eq:DeltaY} or \eqref{eq:Bcomm}.
The \emph{quantum Brascamp--Lieb constant} of the pair $(\mathbf{B},\mathbf{p})$ is
\begin{equation}\label{eq:BL}
f_Q(\mathbf{B},\mathbf{p}) = \sup_{\rho_{XM}\in\mathcal{S}(XM)}\left(S(X|M)(\rho_{XM}) - \sum_{i=1}^K p_i\,S(Y_i|M)(\Phi_{B_i}(\rho_{XM}))\right)\,,
\end{equation}
where $\Phi_{B_i}(\rho_{XM})$ is a shorthand notation for $(\Phi_{B_i}\otimes\mathbb{I}_M)(\rho_{XM})$.
Therefore, any $\rho_{XM}\in\mathcal{S}(XM)$ satisfies the following generalization of the strong subadditivity of the von Neumann entropy:
\begin{equation}\label{eq:QSSA}
S(X|M)(\rho_{XM}) \le \sum_{i=1}^K p_i\,S(Y_i|M)(\Phi_{B_i}(\rho_{XM})) + f_Q(\mathbf{B},\mathbf{p}).
\end{equation}
\end{defn}

The main result of this paper states that, whenever the classical Brascamp--Lieb constant is finite, it coincides with the quantum Brascamp--Lieb constant, and can therefore be computed via the optimization over covariance matrices \eqref{eq:fdet}:
\begin{thm}\label{thm:main}
Let $\mathbf{B}$ be a quantum Brascamp--Lieb datum as in \autoref{defn:QBL} with dimension $(2m,\mathbf{n})$, and let $\mathbf{p}\in\mathbb{R}_{\ge0}^K$ satisfy $2\,m=\mathbf{p}\cdot \mathbf{n}$.
Then, regardless of the dimension of the conditioning quantum system $M$, which can also be chosen trivial, we have
\begin{equation}
f_Q(\mathbf{B},\mathbf{p}) = f(\mathbf{B},\mathbf{p})\,.
\end{equation}
\end{thm}
\autoref{thm:main} has been proved in the particular case without the conditioning quantum system $M$ and when each $Y_i$ is the quantum subsystem of $X$ associated to the Hilbert space $L^2(\mathcal{V}_i)$, where $\mathcal{V}_i$ is a subspace of $\mathbb{R}^m$, and the following condition holds:
\begin{equation}
\sum_{i=1}^K p_i\,\Pi_i = I_m\,,
\end{equation}
where each $\Pi_i$ is the orthogonal projector onto the corresponding $\mathcal{V}_i$ [\onlinecite[Proposition III.4]{berta2019quantum}].
In this case, each $B_i$ selects the quadratures associated with the corresponding $\mathcal{V}_i$ and satisfies $\mathcal{V}_i = \mathrm{Im}\, B_i^T$.

Another result in the same spirit is the generalization of the strong subadditivity of the von Neumann entropy for \emph{fermionic} quantum Gaussian systems proposed and proved in [\onlinecite{carlen2008brascamp}].

The proof of \autoref{thm:main} is based on perturbation with a quantum generalization of the heat semigroup.
In the limit of infinite time, the chosen quantum heat semigroup drives the input state to a quantum Gaussian state with infinite covariance matrix proportional to the covariance matrix of the Gaussian probability distribution that achieves the supremum in \eqref{eq:fdet}.
The technique of proving entropic and functional inequalities via perturbation with the heat semigroup has been first introduced in the classical setting to prove the Entropy Power Inequality [\onlinecite{stam1959some,blachman1965convolution}] and has later been applied to prove Brascamp--Lieb inequalities [\onlinecite{carlen2009subadditivity}].
This technique has been extended to the setting of bosonic quantum systems to prove quantum versions of the Entropy Power Inequality [\onlinecite{konig2014entropy,konig2016corrections,de2014generalization,de2015multimode,koenig2015conditional,de2017gaussian, huber2017geometric,de2018conditional,de2018gaussian,huber2018conditional,de2019entropy,de2019new}] and the geometric quantum Brascamp--Lieb inequality of [\onlinecite[Proposition III.4]{berta2019quantum}].

The paper is structured as follows:
In \autoref{sec:classical}, we collect and put together the results that bring to the expression \eqref{eq:fdet} for the classical Brascamp--Lieb constant and determine whether it is finite.
In \autoref{sec:prel}, we present the required notions and facts on quantum Gaussian systems.
In \autoref{sec:Stam}, we prove a de Bruijn identity, which relates the time derivative of the conditional entropy with respect to the heat semigroup to a Fisher information.
We also prove a Stam inequality stating that the Fisher information is decreasing with respect to $\Phi_B$.
In \autoref{sec:proof}, we employ the Stam inequality to prove that the argument of the supremum in \eqref{eq:BL} increases with time.
Therefore, such supremum is achieved in the limit of infinite time, and its value can be determined from the asymptotic scaling of the conditional entropy with respect to the heat semigroup, which we determine in \autoref{sec:scaling}.

In \autoref{sec:appl}, we present three applications of our results:
\begin{itemize}
  \item In \autoref{sec:EUR}, we discuss the application of the generalized strong subadditivity \eqref{eq:QSSA} as entropic uncertainty relation with quantum memory.
  \item In \autoref{sec:EPI}, we prove a generalization of the quantum conditional Entropy Power Inequality of [\onlinecite{de2018conditional}].
  \item In \autoref{sec:ent}, we determine a lower bound to the bipartite correlations generated by a Gaussian symplectic unitary transformation,
      Moreover, we prove that for a large class of quadratic Hamiltonians, the generated entanglement entropy grows linearly with time for any initial state, settling a conjecture formulated in [\onlinecite{hackl2018entanglement}]\footnote{After the publication of the present paper on arXiv, the generalized strong subadditivity proved in \autoref{thm:main} has been applied to prove the linear scaling of the entanglement entropy for all the unstable quadratic Hamiltonians [\onlinecite{de2022linear}].}.
\end{itemize}

\section{The classical Brascamp--Lieb constant}\label{sec:classical}
In this section, we prove the identity \eqref{eq:fdet} for the classical Brascamp--Lieb constant and determine under which conditions it is finite.

Determining whether the Brascamp--Lieb constant is finite requires the following notion of critical subspace:
\begin{defn}[Critical subspace]
Let $(\mathbf{B},\mathbf{p})$ be a pair where $\mathbf{B}$ is a Brascamp--Lieb datum with dimension $(n,\mathbf{n})$ and $\mathbf{p}\in\mathbb{R}_{\ge0}^K$.
A subspace $\mathcal{V}\subseteq\mathbb{R}^n$ is \emph{subcritical} for the pair $(\mathbf{B},\mathbf{p})$ if
\begin{equation}\label{eq:Vsc}
\dim\mathcal{V} \le \sum_{i=1}^Kp_i\dim B_i\mathcal{V}\,,
\end{equation}
and \emph{critical} if \eqref{eq:Vsc} holds with equality.
\begin{rem}
$\mathbb{R}^n$ is critical iff $n=\mathbf{p}\cdot\mathbf{n}$.
\end{rem}
\end{defn}

The following \autoref{thm:finite-class} provides the condition for the finiteness of the Brascamp--Lieb constant:
\begin{thm}\label{thm:finite-class}
Let $\mathbf{B}$ be a Brascamp--Lieb datum with dimension $(n,\mathbf{n})$.
Then, for any $\mathbf{p}\in\mathbb{R}_{\ge0}^K$, $f(\mathbf{B},\mathbf{p})$ is finite iff $n=\mathbf{p}\cdot\mathbf{n}$ and any subspace of $\mathbb{R}^n$ is subcritical for the pair $(\mathbf{B},\mathbf{p})$.
\end{thm}
\begin{proof}
The claim follows from the equivalence between the finiteness of $f(\mathbf{B},\mathbf{p})$  and the validity of a Brascamp--Lieb inequality with a finite constant [\onlinecite[Theorem 2.1]{carlen2009subadditivity}]  (see \autoref{app:BL}), and the necessary and sufficient condition for the latter [\onlinecite[Theorem 1.13]{bennett2008brascamp}].
\end{proof}

A seminal result by Lieb [\onlinecite[Theorem 6.2]{lieb1990gaussian}] proves that the optimization in \eqref{eq:fX} can be restricted to centered Gaussian random variables, therefore reducing the computation of the Brascamp--Lieb constant to the optimization over covariance matrices given by \eqref{eq:fdet}:
\begin{thm}\label{thm:Lieb}
Let $\mathbf{B}$ be a Brascamp--Lieb datum with dimension $(n,\mathbf{n})$.
Then, for any $\mathbf{p}\in\mathbb{R}_{\ge0}^K$, $f(\mathbf{B},\mathbf{p})$ is given by \eqref{eq:fdet}.
\begin{rem}
We define for any $\alpha\in\mathbb{R}^{n\times n}_{>0}$
\begin{equation}
F_{\mathbf{B},\mathbf{p}}(\alpha) = \frac{1}{2}\ln\det\alpha - \sum_{i=1}^K\frac{p_i}{2}\ln\det\left(B_i\,\alpha\,B_i^T\right)\,,
\end{equation}
such that \eqref{eq:fdet} becomes
\begin{equation}
f(\mathbf{B},\mathbf{p}) = \sup_{\alpha\in\mathbb{R}^{n\times n}_{>0}}F_{\mathbf{B},\mathbf{p}}(\alpha)\,.
\end{equation}
If $f(\mathbf{B},\mathbf{p})$ is finite, from \autoref{thm:finite-class} we have $n = \mathbf{p}\cdot\mathbf{n}$.
In this case, $F_{\mathbf{B},\mathbf{p}}$ is scale-invariant, \emph{i.e.}, for any $\alpha\in\mathbb{R}^{n\times n}_{>0}$ and any $\lambda>0$ we have
\begin{equation}
F_{\mathbf{B},\mathbf{p}}(\lambda\,\alpha) = F_{\mathbf{B},\mathbf{p}}(\alpha)\,.
\end{equation}
Therefore, we refer to the condition $n = \mathbf{p}\cdot\mathbf{n}$ as the \emph{scaling condition}.
\end{rem}
\end{thm}

The following \autoref{thm:class} provides a criterion to determine whether the supremum in \eqref{eq:fdet} is achieved:
\begin{thm}
\label{thm:class}
Let $\mathbf{B}$ be a Brascamp--Lieb datum with dimension $(n,\mathbf{n})$, and let $\mathbf{p}\in\mathbb{R}_{\ge0}^K$.
The following properties are equivalent:
\begin{enumerate}
\item \label{sup=max} The supremum in \eqref{eq:fdet} is achieved.
\item\label{alpha} There exists $\alpha\in\mathbb{R}^{n\times n}_{>0}$ satisfying
\begin{equation}\label{eq:condalpha}
\sum_{i=1}^K p_i\,B_i^T\left(B_i\,\alpha\,B_i^T\right)^{-1}B_i = \alpha^{-1}\,.
\end{equation}
\item\label{critical_complementary} The scaling condition $n=\mathbf{p}\cdot\mathbf{n}$ holds, any subspace of $\mathbb{R}^n$ is subcritical for the pair $(\mathbf{B},\mathbf{p})$, and any critical subspace of $\mathbb{R}^n$ has a critical complementary subspace.
\end{enumerate}
Moreover, if any of the above condition holds, the $\alpha\in\mathbb{R}^{n\times n}_{>0}$ satisfying \eqref{eq:condalpha} achieves the supremum in \eqref{eq:fdet}.
\end{thm}
\begin{proof}
\begin{description}
\item[\ref{sup=max} $\Rightarrow$ \ref{alpha}]
Let $\alpha\in\mathbb{R}^{n\times n}_{>0}$ achieve the global maximum of $F_{\mathbf{B},\mathbf{p}}$.
In particular, $\alpha$ is a stationary point of $F_{\mathbf{B},\mathbf{p}}$.
We have
\begin{equation}
\mathrm{d}F_{\mathbf{B},\mathbf{p}}(\alpha) = \frac{1}{2}\,\mathrm{tr}\left[\mathrm{d}\alpha\left(\alpha^{-1} - \sum_{i=1}^Kp_i\,B_i^T\left(B_i\,\alpha\,B_i^T\right)^{-1}B_i\right)\right]\,,
\end{equation}
and \eqref{eq:condalpha} follows.
\item[\ref{alpha} $\Leftrightarrow$ \ref{critical_complementary}]
corresponds to [\onlinecite[Theorem 7.13, (d) $\Leftrightarrow$ (h)]{bennett2008brascamp}] letting $A_i =  \left(B_i\,\alpha\,B_i^T\right)^{-1}$.
\item[\ref{alpha} $\Rightarrow$ \ref{sup=max}]
follows from [\onlinecite[Proposition 3.6, (c) $\Rightarrow$ (a)]{bennett2008brascamp}], yielding that Gaussian extremizers exist for the dual Brascamp--Lieb inequality, and the correspondence between optimizers in the entropy inequality [\onlinecite[Theorem 2.2]{carlen2009subadditivity}]. Following the construction of such optimizers in fact proves that $\alpha$ satisfying \eqref{eq:condalpha} achieves the supremum in \eqref{eq:fdet}.
\end{description}
\end{proof}

\section{Preliminaries}\label{sec:prel}
This section contains some notation and facts on quantum Gaussian systems that are needed in the proofs.
For further details, we refer the reader to the books [\onlinecite{ferraro2005gaussian}], [\onlinecite{serafini2017quantum}] or [\onlinecite[Chapter 12]{holevo2019quantum}].
In order to avoid having separate proofs in the cases where $Y_i$ is a quantum Gaussian system or a random variable, at the price of some abuse we employ wherever possible the same notation for the classical and quantum case.

Let $X$ be a quantum Gaussian system with $m$ modes.
For any $x\in\mathbb{R}^{2m}$, the \emph{displacement operator} associated to $x$ is the unitary operator $D_x^X$ acting on $L^2(\mathbb{R}^m)$ such that
\begin{equation}
D^{X\dag}_x\,R_i^X\,D^X_x = R_i^X + x_i\,\mathbb{I}
\end{equation}
for any $i=1,\,\ldots,\,2m$.

With some abuse of notation, for any random variable $X$ with values in $\mathbb{R}^n$ and probability distribution $\mu_X$ and for any $x\in\mathbb{R}^n$ we denote with $D_x^X\,\mu_X\,D^{X\dag}_x$ the probability distribution of $X + x$.

Let $\rho_X$ be a quantum state of $X$ with finite energy.
The vector of the first moments of $\rho_X$ is
\begin{equation}
r_i = \mathrm{Tr}\left[\rho_X\,R_i^X\right]\,,\qquad i=1,\,\ldots,\,2m\,,
\end{equation}
and the covariance matrix of $\rho_X$ is
\begin{align}
&\gamma_{ij} = \frac{1}{2}\,\mathrm{Tr}\left[\left(R_i^X - r_i\,\mathbb{I}\right)\rho_X\left(R_j^X - r_j\,\mathbb{I}\right) + \left(R_j^X - r_j\,\mathbb{I}\right)\rho_X\left(R_i^X - r_i\,\mathbb{I}\right)\right]\,,\nonumber\\
&i,\,j=1,\,\ldots,\,2m\,.
\end{align}
The symplectic eigenvalues of a matrix $\gamma\in\mathbb{R}^{2m\times2m}_{>0}$ are the absolute values of the eigenvalues of $\gamma\,\Delta_{2m}^{-1}$, and their multiplicity is always even.
We denote with $\nu_{\min}(\gamma)$ the minimum symplectic eigenvalue of $\gamma$.
If $\gamma$ is the covariance matrix of a quantum state, it satisfies $\nu_{\min}(\gamma)\ge1/2$.

A Gaussian state of $X$ is a quantum state that is either proportional to the exponential of a Hamiltonian given by a quadratic polynomial in the quadratures or is the unique ground state of such a Hamiltonian.
A Gaussian state is uniquely determined by its covariance matrix and by the vector of its first moments.
For any $\gamma\in\mathbb{R}^{2m\times2m}_{>0}$ with $\nu_{\min}(\gamma) \ge 1/2$, we denote with $\omega_X(\gamma)$ the centered quantum Gaussian state of $X$ with covariance matrix $\gamma$.
Quantum Gaussian states maximize the von Neumann entropy among all the quantum states with a given covariance matrix, \emph{i.e.}, if $\rho_X$ is a quantum state of $X$ with covariance matrix $\gamma$, then $S(X)(\rho_X) \le S(X)(\omega_X(\gamma))$.

Analogously, if $X$ is a random variable with values in $\mathbb{R}^n$, for any $\gamma\in\mathbb{R}^{n\times n}_{>0}$ we denote with $\omega_X(\gamma)$ the centered Gaussian probability distribution of $X$ with covariance matrix $\gamma$.

\begin{defn}[Heat semigroup]
Let $\alpha\in\mathbb{R}^{n\times n}_{\ge0}$ (the set of the symmetric and positive semidefinite $n\times n$ real matrices), and let $Z$ be a centered Gaussian random variables with values in $\mathbb{R}^n$ and covariance matrix $\alpha$.

The \emph{classical heat semigroup} with covariance matrix $\alpha$ is the stochastic map $\mathcal{N}^X_\alpha$ that sends the probability distribution of a random variable $X$ with values in $\mathbb{R}^n$ to the probability distribution of $X+Z$.

For even $n$, the \emph{quantum heat semigroup} with covariance matrix $\alpha$ is the quantum channel $\mathcal{N}^X_\alpha$ acting on a quantum state $\rho_X$ of the quantum Gaussian system $X$ with $n/2$ modes as
\begin{equation}
\mathcal{N}^X_\alpha(\rho_X) = \mathbb{E}_Z\left(D^X_Z\,\rho_X\,D^{X\dag}_Z\right)\,.
\end{equation}

Both the classical and quantum heat semigroups satisfy the semigroup property $\mathcal{N}^X_\alpha\circ\mathcal{N}^X_\beta = \mathcal{N}^X_{\alpha+\beta}$ for any $\alpha,\,\beta\in\mathbb{R}^{n\times n}_{\ge0}$ and commute with displacements.
\end{defn}

Let $B\in\mathbb{R}^{n\times2m}$ satisfy either \eqref{eq:DeltaY} or \eqref{eq:Bcomm}.
Then, the heat semigroup and the displacement operators have the following commutation properties with $\Phi_B$:
\begin{lem}\label{lem:NPhi}
For any $\alpha\in\mathbb{R}^{2m\times2m}_{\ge0}$ and any $x\in\mathbb{R}^{2m}$,
\begin{equation}
\Phi_{B}\circ\mathcal{N}^X_\alpha = \mathcal{N}^{Y}_{B\alpha B^T}\circ\Phi_{B}\,,\qquad \Phi_{B}\left(D^X_x\,\cdot\,D^{X\dag}_x\right) = D^{Y}_{Bx}\,\Phi_{B}(\,\cdot\,)\,D^{Y\dag}_{Bx}\,.
\end{equation}
\end{lem}

\section{Stam inequality}\label{sec:Stam}
In this section, we prove the de Bruijn identity \autoref{prop:dBJ}, which relates the time derivative of the conditional entropy with respect to the heat semigroup to the Fisher information with respect to the displacements.
We also prove the Stam inequality \autoref{prop:Stam}, stating that the Fisher information is decreasing with respect to $\Phi_B$.

All the results cover both the cases where $X$ is a quantum Gaussian system with $n/2$ modes (with $n$ even) or a random variable with values in $\mathbb{R}^n$; $M$ will always be a conditioning quantum system.
In the case where $X$ is a random variable, we denote with $\mathcal{S}(XM)$ the set of the joint states of $XM$ such that $S(M)$ is finite and $X$ has finite energy, \emph{i.e.}, if $p$ is the probability density of $X$,
\begin{equation}\label{eq:enX}
\mathbb{E}_X\left(\left|X\right|^2 + \frac{1}{4}\left|\nabla\ln p(X)\right|^2\right) < \infty\,.
\end{equation}
Condition \eqref{eq:enX} is always satisfied by the random variable associated to the outcome of the joint measurement of a set of commuting quadratures on a quantum state with finite energy (\autoref{lem:finite-energy-classical} of \autoref{app:entropy}) and entails that $X$ has finite entropy (\autoref{lem:finite-entropy-classical} of \autoref{app:entropy}).

To avoid regularity issues, we first define the Fisher information in the following integral form:
\begin{defn}[Integral Fisher information]\label{defn:Delta}
For any $\rho_{XM}\in\mathcal{S}(XM)$ and any $\alpha\in\mathbb{R}^{n\times n}_{\ge0}$, we define
\begin{equation}
\Delta_{X|M}(\rho_{XM})(\alpha) = I(X;Z|M)(\sigma_{XMZ}) \ge 0\,,
\end{equation}
where
\begin{equation}
I(X;Z|M) = S(X|M) + S(Z|M) - S(XZ|M)
\end{equation}
is the conditional mutual information and the state $\sigma_{XMZ}$ is defined as follows:
$Z$ is a centered Gaussian random variable with values in $\mathbb{R}^{n}$ and covariance matrix $\alpha$, and for any $z\in\mathbb{R}^n$, the state of $XM$ conditioned on $Z=z$ is
\begin{equation}
\sigma_{XM}(z) = D^X_z\,\rho_{XM}\,D^{X\dag}_z\,,
\end{equation}
such that
\begin{equation}
\sigma_{XM} = \mathcal{N}^X_\alpha(\rho_{XM})\,.
\end{equation}
\end{defn}

The integral Fisher information is decreasing with respect to the heat semigroup:
\begin{prop}\label{prop:Deltadp}
For any $\rho_{XM}\in\mathcal{S}(XM)$ and any $\alpha,\,\beta\in\mathbb{R}^{n\times n}_{\ge0}$,
\begin{equation}
\Delta_{X|M}\left(\mathcal{N}^X_\beta(\rho_{XM})\right)(\alpha) = I(X;Z|M)\left(\mathcal{N}^X_\beta(\sigma_{XMZ})\right) \le \Delta_{X|M}(\rho_{XM})(\alpha)\,,
\end{equation}
where $\sigma_{XMZ}$ is as in \autoref{defn:Delta}.
\end{prop} \begin{proof}
We have
\begin{equation}
\Delta_{X|M}\left(\mathcal{N}^X_\beta(\rho_{XM})\right)(\alpha) = I(X;Z|M)(\tau_{XMZ})\,,
\end{equation}
where the state $\tau_{XMZ}$ is defined as follows:
$Z$ is a centered Gaussian random variable with values in $\mathbb{R}^n$ and covariance matrix $\alpha$, and for any $z\in\mathbb{R}^n$, the state of $XM$ conditioned on $Z=z$ is
\begin{equation}
\tau_{XM}(z) = D^X_z\,\mathcal{N}^X_\beta(\rho_{XM})\,D^{X\dag}_z = \mathcal{N}^X_\beta\left(D^X_z\,\rho_{XM}\,D^{X\dag}_z\right) = \mathcal{N}^X_\beta(\sigma_{XM}(z))\,.
\end{equation}
We then have
\begin{equation}
\tau_{XMZ} = \mathcal{N}^X_\beta(\sigma_{XMZ})\,,
\end{equation}
and the data-processing inequality for the conditional mutual information implies
\begin{align}
\Delta_{X|M}\left(\mathcal{N}^X_\beta(\rho_{XM})\right)(\alpha) &= I(X;Z|M)\left(\mathcal{N}^X_\beta(\sigma_{XMZ})\right) \le I(X;Z|M)(\sigma_{XMZ})\nonumber\\
&= \Delta_{X|M}(\rho_{XM})(\alpha)\,.
\end{align}
The claim follows.
\end{proof}

The following integral de Bruijn identity connects the integral Fisher information with the heat semigroup by proving that the former is equal to the increase in the conditional entropy induced by the latter.
The de Bruijn identity \autoref{prop:dBJ} will follow from this integral version by suitably taking the limit $\alpha\to0$.
\begin{prop}[Integral de Bruijn identity]\label{prop:dB}
For any $\rho_{XM}\in\mathcal{S}(XM)$ and any $\alpha\in\mathbb{R}^{n\times n}_{\ge0}$,
\begin{equation}
\Delta_{X|M}(\rho_{XM})(\alpha) = S(X|M)\left(\mathcal{N}^X_\alpha(\rho_{XM})\right) - S(X|M)(\rho_{XM})\,.
\end{equation}
\end{prop} \begin{proof}
Let $\sigma_{XMZ}$ be as in \autoref{defn:Delta}.
We have
\begin{align}
\Delta_{X|M}(\rho_{XM})(\alpha) &= I(X;Z|M)(\sigma_{XMZ}) = S(X|M)(\sigma_{XM}) - S(X|MZ)(\sigma_{XMZ})\nonumber\\
&= S(X|M)\left(\mathcal{N}^X_\alpha(\rho_{XM})\right) - \mathbb{E}_Z\,S(X|M)\left(D^X_Z\,\rho_{XM}\,D^{X\dag}_Z\right)\nonumber\\
&= S(X|M)\left(\mathcal{N}^X_\alpha(\rho_{XM})\right) - S(X|M)(\rho_{XM})\,,
\end{align}
and the claim follows.
\end{proof}

We will now prove the properties of the integral Fisher information that will be needed to define the Fisher information.

\begin{itemize}
\item The integral Fisher information is an increasing and subadditive function of the covariance matrix:
\begin{prop}\label{prop:sub}
For any $\rho_{XM}\in\mathcal{S}(XM)$, the function
\begin{equation}\label{eq:falpha}
\alpha \mapsto \Delta_{X|M}(\rho_{XM})(\alpha):\alpha\in\mathbb{R}^{n\times n}_{\ge0}
\end{equation}
is increasing and subadditive.
\end{prop} \begin{proof}
Let $\alpha,\,\beta\in\mathbb{R}^{n\times n}_{\ge0}$.
We have from \autoref{prop:dB}
\begin{align}
\Delta_{X|M}(\rho_{XM})(\alpha + \beta) &= S(X|M)\left(\mathcal{N}^X_{\alpha+\beta}(\rho_{XM})\right) - S(X|M)(\rho_{XM})\nonumber\\
&= S(X|M)\left(\mathcal{N}^X_\alpha\left(\mathcal{N}^X_\beta(\rho_{XM})\right)\right) - S(X|M)\left(\mathcal{N}^X_\beta(\rho_{XM})\right)\nonumber\\
&\phantom{=}\, + S(X|M)\left(\mathcal{N}^X_\beta(\rho_{XM})\right) - S(X|M)(\rho_{XM})\nonumber\\
&= \Delta_{X|M}\left(\mathcal{N}^X_\beta(\rho_{XM})\right)(\alpha) + \Delta_{X|M}(\rho_{XM})(\beta) \ge \Delta_{X|M}(\rho_{XM})(\beta)\,,
\end{align}
therefore the function \eqref{eq:falpha} is increasing.
We have from \autoref{prop:Deltadp}
\begin{align}
\Delta_{X|M}(\rho_{XM})(\alpha + \beta) &= \Delta_{X|M}\left(\mathcal{N}^X_\beta(\rho_{XM})\right)(\alpha) + \Delta_{X|M}(\rho_{XM})(\beta)\nonumber\\
&\le \Delta_{X|M}(\rho_{XM})(\alpha) + \Delta_{X|M}(\rho_{XM})(\beta)\,,
\end{align}
therefore the function \eqref{eq:falpha} is subadditive.
The claim follows.
\end{proof}

\item The integral Fisher information as a function of the covariance matrix satisfies the following weaker version of concavity:
\begin{prop}\label{prop:matconc}
For any $\rho_{XM}\in\mathcal{S}(XM)$ and any $\alpha,\,\beta\in\mathbb{R}^{n\times n}_{\ge0}$ we have
\begin{equation}
2\,\Delta_{X|M}(\rho_{XM})(\alpha+\beta) \ge \Delta_{X|M}(\rho_{XM})(\alpha) + \Delta_{X|M}(\rho_{XM})(\alpha+2\,\beta)\,.
\end{equation}
\end{prop} \begin{proof}
We have with the help of \autoref{prop:dB}
\begin{align}
&2\,\Delta_{X|M}(\rho_{XM})(\alpha + \beta) - \Delta_{X|M}(\rho_{XM})(\alpha) - \Delta_{X|M}(\rho_{XM})(\alpha + 2\,\beta)\nonumber\\
&= 2\,S(X|M)\left(\mathcal{N}^X_{\alpha+\beta}(\rho_{XM})\right) - S(X|M)\left(\mathcal{N}^X_{\alpha}(\rho_{XM})\right) - S(X|M)\left(\mathcal{N}^X_{\alpha+2\beta}(\rho_{XM})\right)\nonumber\\
&=S(X|M)\left(\mathcal{N}^X_{\beta}\left(\mathcal{N}^X_{\alpha}(\rho_{XM})\right)\right) - S(X|M)\left(\mathcal{N}^X_{\alpha}(\rho_{XM})\right)\nonumber\\
& \phantom{=} - S(X|M)\left(\mathcal{N}^X_{\beta}\left(\mathcal{N}^X_{\alpha+\beta}(\rho_{XM})\right)\right) + S(X|M)\left(\mathcal{N}^X_{\alpha+\beta}(\rho_{XM})\right)\nonumber\\
&= \Delta_{X|M}\left(\mathcal{N}^X_{\alpha}(\rho_{XM})\right)(\beta) - \Delta_{X|M}\left(\mathcal{N}^X_{\alpha+\beta}(\rho_{XM})\right)(\beta)\nonumber\\
&= \Delta_{X|M}\left(\mathcal{N}^X_{\alpha}(\rho_{XM})\right)(\beta) - \Delta_{X|M}\left(\mathcal{N}^X_{\beta}\left(\mathcal{N}^X_{\alpha}(\rho_{XM})\right)\right)(\beta) \ge 0\,,
\end{align}
where the last inequality follows from \autoref{prop:Deltadp}.
The claim follows.
\end{proof}

\item The integral Fisher information is a continuous, increasing and concave function of the covariance matrix if restricted to a half-line starting from the origin:
\begin{prop}\label{prop:tconcave}
For any $\rho_{XM}\in\mathcal{S}(XM)$ and any $\alpha\in\mathbb{R}^{n\times n}_{\ge0}$, the function
\begin{equation}\label{eq:ft}
t\mapsto \Delta_{X|M}(\rho_{XM})(t\,\alpha)\,,\qquad t\ge0
\end{equation}
is continuous, increasing and concave.
\end{prop} \begin{proof}
The function \eqref{eq:ft} is increasing for \autoref{prop:sub}.
Let $0\le s\le t$.
We have with the help of \autoref{prop:matconc}
\begin{align}
2\,\Delta_{X|M}(\rho_{XM})\left(\tfrac{t+s}{2}\,\alpha\right) &= 2\,\Delta_{X|M}(\rho_{XM})\left(s\,\alpha + \tfrac{t-s}{2}\,\alpha\right)\nonumber\\
&\ge \Delta_{X|M}(\rho_{XM})(s\,\alpha) + \Delta_{X|M}(\rho_{XM})\left(s\,\alpha + \left(t-s\right)\alpha\right)\nonumber\\
&= \Delta_{X|M}(\rho_{XM})(s\,\alpha) + \Delta_{X|M}(\rho_{XM})(t\,\alpha)\,,
\end{align}
therefore the function \eqref{eq:ft} is midpoint concave and therefore concave.
Then, the function \eqref{eq:ft} is continuous for any $t>0$ and lower semicontinuous in $t=0$.

 We have from \autoref{prop:dB}
\begin{equation}
\Delta_{X|M}(\rho_{XM})(t\,\alpha) = S(X|M)\left(\mathcal{N}^X_{t\alpha}(\rho_{XM})\right) - S(X|M)(\rho_{XM})\,,
\end{equation}
therefore the continuity of the function \eqref{eq:ft} in $t=0$ follows if we prove that the function
\begin{equation}
t\mapsto S(X|M)\left(\mathcal{N}^X_{t\alpha}(\rho_{XM})\right)
\end{equation}
 is upper semicontinuous in $t=0$.
\begin{description}
\item[Quantum case] If $X$ is a quantum Gaussian system with $n/2$ modes, let
\begin{equation}
H_X = \sum_{i=1}^{n}\left(R_i^X\right)^2\,, \qquad \omega_X = \frac{\mathrm{e}^{-H_X}}{\mathrm{Tr}\,\mathrm{e}^{-H_X}}\,.
\end{equation}
We denote with
\begin{equation}
S(\rho\|\sigma) = \mathrm{Tr}\left[\rho\left(\ln\rho - \ln\sigma\right)\right]
\end{equation}
the relative entropy between the quantum states $\rho$ and $\sigma$ [\onlinecite{holevo2019quantum}].
We have for any $t\ge0$
\begin{align}\label{eq:SR}
S(X|M)\left(\mathcal{N}^X_{t\alpha}(\rho_{XM})\right) &= \sum_{i=1}^n\mathrm{Tr}\left[R^X_i\,\mathcal{N}^X_{t\alpha}(\rho_{X})\,R^X_i\right] + \ln\mathrm{Tr}\,\mathrm{e}^{-H_X}\nonumber\\
&\phantom{=} \, - S\left(\left.\mathcal{N}^X_{t\alpha}(\rho_{XM})\right\|\omega_X\otimes\rho_M\right)\,.
\end{align}
The function $t\mapsto\mathcal{N}^X_{t\alpha}(\rho_{XM})$ is continuous with respect to the trace norm and the relative entropy is a lower semicontinuous function of the first argument [\onlinecite[Theorem 11.6]{holevo2019quantum}].
Therefore, the function $t\mapsto S\left(\left.\mathcal{N}^X_{t\alpha}(\rho_{XM})\right\|\omega_X\otimes\rho_M\right)$ is lower semicontinuous.
Since the function $t\mapsto\mathrm{Tr}\left[R^X_i\,\mathcal{N}^X_{t\alpha}(\rho_{X})\,R^X_i\right]$ is affine for any $i=1,\,\ldots,\,n$, from \eqref{eq:SR} the function $t\mapsto S(X|M)\left(\mathcal{N}^X_{t\alpha}(\rho_{XM})\right)$ is upper semicontinuous.

\item[Classical case] If $X$ is a random variable with values in $\mathbb{R}^n$ and
\begin{equation}
\rho_{XM} = \left\{\rho_M(x):x\in\mathbb{R}^n\right\}
\end{equation}
is a joint state of $XM$, then
\begin{equation}
\mathcal{N}^X_{t\alpha}(\rho_{XM}) = \left \{ \mathbb{E}_Z\,\rho_{M}\left( x+ \sqrt{t}\, Z\right): x \in \mathbb{R}^n \right\}\, ,
\end{equation}
 where $Z$ is a centered Gaussian random variable with values in $\mathbb{R}^n$ and covariance matrix $\alpha$ .
The properties of the convolution give that
 \begin{equation}
\lim_{t\to0}\mathbb{E}_X\left\|\mathbb{E}_Z\,\rho_{M}\left( X+ \sqrt{t} \,Z\right) - \rho_M(X)\right\|_1 = 0\,,
\end{equation}
where $\left\|\cdot\right\|_1$ denotes the trace norm.
We recall that an equivalent definition of the conditional entropy is
\begin{equation}\label{eq:conditional-entropy-classical-quantum}
 S(X|M)(\rho_{XM}) = S(X) - I(X; M)\, ,
 \end{equation}
 where the mutual information is defined in terms of the relative entropy
 \begin{equation}\label{eq:mutual-info-general}
 I(X; M )(\rho_{XM}) = S( \rho_{XM}  \| \rho_X \otimes \rho_M )\,,
 \end{equation}
which can be defined in the context of general von Neumann algebras [\onlinecite{ohya2004quantum}].
The relative entropy is lower semicontinuous with respect to the weak convergence of the states [\onlinecite[Corollary 5.12]{ohya2004quantum}]. Therefore, by \eqref{eq:mutual-info-general}, the function $t \mapsto I(X;M)\left(\mathcal{N}^X_{t\alpha}(\rho_{XM}) \right)$ is lower semicontinuous.
Let $\omega_X$ be the Gaussian probability distribution for $X$ with density
\begin{equation}
\omega_X(\mathrm{d}^nx) = \mathrm{e}^{-\frac{|x|^2}{2}}\,\frac{\mathrm{d}^nx}{\left(2\pi\right)^\frac{n}{2}}\, ,\qquad x\in\mathbb{R}^n\,.
\end{equation}
Then, also the function $t \mapsto S\left(\left.\mathcal{N}^X_{t\alpha}(\rho_{X}) \right\| \omega_X\right)$ is lower semicontinuous and, writing
\begin{equation}
S(X)\left(\mathcal{N}^X_{t\alpha}(\rho_{X})\right) = \frac{1}{2}\,\mathbb{E}_{XZ} \left|X+ \sqrt t \,Z\right|^2 - S\left(\left.\mathcal{N}^X_{t\alpha}(\rho_{X}) \right\| \omega_X\right)\,,
\end{equation}
the function $t \mapsto S(X)\left(\mathcal{N}^X_{t\alpha}(\rho_{X})\right)$ is upper semicontinuous.
By \eqref{eq:conditional-entropy-classical-quantum}, we conclude that the function $t\mapsto S(X|M)\left(\mathcal{N}^X_{t\alpha}(\rho_{XM})\right)$ is upper semicontinuous.
\end{description}
The claim follows.
\end{proof}
\end{itemize}

We can now define the Fisher information:
\begin{defn}[Fisher information]
For any $\rho_{XM}\in\mathcal{S}(XM)$ and any $\alpha\in\mathbb{R}^{n\times n}_{\ge0}$, we define
\begin{equation}\label{eq:defJ}
J_{X|M}(\rho_{XM})(\alpha) = \lim_{t\to0}\frac{\Delta_{X|M}(\rho_{XM})(t\,\alpha)}{t}\,.
\end{equation}
\end{defn}

\begin{rem}
From \autoref{prop:tconcave}, the limit in \eqref{eq:defJ} always exists, finite or infinite.
\end{rem}

We aim to prove that the Fisher information is a linear function of the covariance matrix.
We need the following intermediate results:
\begin{lem}\label{lem:Jt}
For any $\rho_{XM}\in\mathcal{S}(XM)$, any $\alpha\in\mathbb{R}^{n\times n}_{\ge0}$ and any $t\ge0$ we have
\begin{equation}
J_{X|M}(\rho_{XM})(t\,\alpha) = t\,J_{X|M}(\rho_{XM})(\alpha)\,.
\end{equation}
\end{lem}
\begin{proof}
We have
\begin{align}
J_{X|M}(\rho_{XM})(t\,\alpha) &= \lim_{s\to0}\frac{\Delta_{X|M}(\rho_{XM})(s\,t\,\alpha)}{s} = t\lim_{s\to0}\frac{\Delta_{X|M}(\rho_{XM})(s\,\alpha)}{s}\nonumber\\
&= t\,J_{X|M}(\rho_{XM})(\alpha)\,,
\end{align}
and the claim follows.
\end{proof}

\begin{lem}\label{lem:Jsub}
For any $\rho_{XM}\in\mathcal{S}(XM)$, the function
\begin{equation}
\alpha \mapsto J_{X|M}(\rho_{XM})(\alpha):\alpha\in\mathbb{R}^{n\times n}_{\ge0}
\end{equation}
is increasing and subadditive.
\end{lem}
\begin{proof}
Follows from \autoref{prop:sub}.
\end{proof}

\begin{lem}\label{lem:Jconc}
For any $\rho_{XM}\in\mathcal{S}(XM)$ and any $\alpha,\,\beta\in\mathbb{R}^{n\times n}_{\ge0}$ we have
\begin{equation}
2\,J_{X|M}(\rho_{XM})(\alpha+\beta) \ge J_{X|M}(\rho_{XM})(\alpha) + J_{X|M}(\rho_{XM})(\alpha+2\,\beta)\,.
\end{equation}
\end{lem}
\begin{proof}
Follows from \autoref{prop:matconc}.
\end{proof}

\begin{lem}\label{lem:Jaddp}
For any $\rho_{XM}\in\mathcal{S}(XM)$ and any $\alpha,\,\beta\in\mathbb{R}^{n\times n}_{\ge0}$ we have
\begin{equation}
J_{X|M}(\rho_{XM})(2\,\alpha+\beta) = J_{X|M}(\rho_{XM})(\alpha) + J_{X|M}(\rho_{XM})(\alpha + \beta)\,.
\end{equation}
\end{lem}
\begin{proof}
We have
\begin{align}
J_{X|M}(\rho_{XM})(\alpha) + J_{X|M}(\rho_{XM})(\alpha + \beta) &\overset{\mathrm{(a)}}{\le} 2\,J_{X|M}(\rho_{XM})\left(\alpha+\frac{\beta}{2}\right)\nonumber\\
&\overset{\mathrm{(b)}}{=} J_{X|M}(\rho_{XM})(\alpha+(\alpha+\beta))\nonumber\\
& \overset{\mathrm{(c)}}{\le} J_{X|M}(\rho_{XM})(\alpha) + J_{X|M}(\rho_{XM})(\alpha + \beta)\,,
\end{align}
where (a) follows from \autoref{lem:Jconc}, (b) from \autoref{lem:Jt} and (c) from \autoref{lem:Jsub}.
The claim follows.
\end{proof}

\begin{prop}\label{prop:linearJ}
For any $\rho_{XM}\in\mathcal{S}(XM)$, the function
\begin{equation}
\alpha \mapsto J_{X|M}(\rho_{XM})(\alpha):\alpha\in\mathbb{R}^{n\times n}_{\ge0}
\end{equation}
is linear.
\end{prop} \begin{proof}
Let $\alpha,\,\beta\in\mathbb{R}^{n\times n}_{\ge0}$.
We have
\begin{align}
2\,J_{X|M}(\rho_{XM})(\alpha+\beta) &\overset{\mathrm{(a)}}{=} J_{X|M}(\rho_{XM})(2\,\alpha + 2\,\beta)\nonumber\\
&\overset{\mathrm{(b)}}{=} J_{X|M}(\rho_{XM})(\alpha) + J_{X|M}(\rho_{XM})(\alpha + 2\,\beta)\nonumber\\
&\overset{\mathrm{(c)}}{=} J_{X|M}(\rho_{XM})(\alpha) + J_{X|M}(\rho_{XM})(\beta) + J_{X|M}(\rho_{XM})(\alpha+\beta)\,,
\end{align}
where (a) follows from \autoref{lem:Jt}, and (b) and (c) follow from \autoref{lem:Jaddp}.
Therefore,
\begin{equation}
J_{X|M}(\rho_{XM})(\alpha+\beta) = J_{X|M}(\rho_{XM})(\alpha) + J_{X|M}(\rho_{XM})(\beta)\,,
\end{equation}
and the claim follows.
\end{proof}

The following de Bruijn identity relates the Fisher information to the time derivative of the conditional entropy with respect to the heat semigroup:
\begin{prop}[De Bruijn identity]\label{prop:dBJ}
For any $\rho_{XM}\in\mathcal{S}(XM)$ and any $\alpha\in\mathbb{R}^{n\times n}_{\ge0}$,
\begin{equation}
J_{X|M}(\rho_{XM})(\alpha) = \left.\frac{\mathrm{d}}{\mathrm{d}t}S(X|M)\left(\mathcal{N}^X_{t\alpha}(\rho_{XM})\right)\right|_{t=0}\,.
\end{equation}
\end{prop} \begin{proof}
Follows from \autoref{prop:dB}.
\end{proof}

The following integral Stam inequality proves that the integral Fisher information is decreasing with respect to the application of $\Phi_B$:
\begin{prop}[Integral Stam inequality]\label{prop:iStam}
We have for any $\rho_{XM}\in\mathcal{S}(XM)$, any $\alpha\in\mathbb{R}^{2m\times2m}_{\ge0}$ and any $B\in\mathbb{R}^{n\times2m}$ satisfying either \eqref{eq:DeltaY} or \eqref{eq:Bcomm}
\begin{equation}
\Delta_{X|M}(\rho_{XM})(\alpha) \ge \Delta_{Y|M}(\Phi_{B}(\rho_{XM}))\left(B\,\alpha\,B^T\right)\,.
\end{equation}
\end{prop} \begin{proof}
Let $\sigma_{XMZ}$ be as in \autoref{defn:Delta}.
We have from the data-processing inequality for the conditional mutual information
\begin{equation}
\Delta_{X|M}(\rho_{XM})(\alpha) = I(X;Z|M)(\sigma_{XMZ}) \ge I(Y;Z|M)(\Phi_{B}(\sigma_{XMZ}))\,.
\end{equation}
We have for any $z\in\mathbb{R}^{2m}$
\begin{equation}
\Phi_{B}(\sigma_{XM}(z)) = \Phi_{B}\left(D^X_z\,\rho_{XM}\,D^{X\dag}_z\right) = D^{Y}_{Bz}\,\Phi_{B}(\rho_{XM})\,D^{Y\dag}_{Bz}\,,
\end{equation}
and $BZ$ is a centered Gaussian random variable with covariance matrix $B\,\alpha\,B^T$, therefore
\begin{equation}
I(Y;Z|M)(\Phi_{B}(\sigma_{XMZ})) = \Delta_{Y|M}(\Phi_{B}(\rho_{XM}))\left(B\,\alpha\,B^T\right)\,.
\end{equation}
The claim follows.
\end{proof}

We can finally prove the following Stam inequality:
\begin{prop}[Stam inequality]\label{prop:Stam}
We have for any $\rho_{XM}\in\mathcal{S}(XM)$, any $B\in\mathbb{R}^{n\times2m}$ satisfying either \eqref{eq:DeltaY} or \eqref{eq:Bcomm} and any $\alpha\in\mathbb{R}^{2m\times2m}_{\ge0}$
\begin{equation}
J_{X|M}(\rho_{XM})(\alpha) \ge J_{Y|M}(\Phi_{B}(\rho_{XM}))\left(B\,\alpha\,B^T\right)\,.
\end{equation}
\end{prop} \begin{proof}
Follows from \autoref{prop:iStam}.
\end{proof}

\section{Asymptotic scaling of the conditional entropy}\label{sec:scaling}
In this section, we determine the asymptotic scaling of the conditional entropy with respect to the time evolution induced by a heat semigroup with a positive definite covariance matrix.
\begin{prop}[Asymptotic scaling of the entropy]\label{prop:as}
Let $X$ be either a quantum Gaussian system with $n/2$ modes or a random variable with values in $\mathbb{R}^n$.
Then, for any $\rho_{XM}\in\mathcal{S}(XM)$ and any $\alpha\in\mathbb{R}^{n\times n}_{>0}$, for $t\to\infty$
\begin{equation}\label{eq:scaling}
S(X|M)\left(\mathcal{N}^X_{t\alpha}(\rho_{XM})\right) = \frac{1}{2}\ln\det\left(\mathrm{e}\,t\,\alpha\right) + O\left(\frac{1}{t}\right)\,.
\end{equation}
\begin{rem}
The right-hand side of \eqref{eq:scaling} does not depend on the initial state $\rho_{XM}$.
\end{rem}
\end{prop} \begin{proof}
For any $\gamma\in\mathbb{R}^{n\times n}_{>0}$, we denote with
\begin{equation}
S_G(\gamma) = \frac{1}{2}\ln\det\left(\mathrm{e}\,\gamma\right)
\end{equation}
the Shannon differential entropy of a Gaussian random variable with covariance matrix $\gamma$.
If $n$ is even and $\nu_{\min}(\gamma)\ge1/2$, we denote with $S_Q(\gamma)$ the von Neumann entropy of the quantum Gaussian state with covariance matrix $\gamma$.

On the one hand, we have for the subadditivity of the entropy
\begin{equation}
S(X|M)\left(\mathcal{N}^X_{t\alpha}(\rho_{XM})\right) \le S(X)\left(\mathcal{N}^X_{t\alpha}(\rho_{X})\right)\,.
\end{equation}
Let $\gamma\in\mathbb{R}^{n\times n}$ be the covariance matrix of $\rho_X$.
$\mathcal{N}^X_{t\alpha}(\rho_{X})$ has covariance matrix $\gamma + t\,\alpha$, and we have from \autoref{lem:SGas} of \autoref{app:SG}
\begin{equation}
S(X)\left(\mathcal{N}^X_{t\alpha}(\rho_{X})\right) \le S(X)(\omega_X(\gamma+t\,\alpha)) = \frac{1}{2}\ln\det\left(\mathrm{e}\,t\,\alpha\right) + O\left(\frac{1}{t}\right)\,.
\end{equation}

The proof of the converse inequality is different for the classical and the quantum case:
\begin{description}
\item[Quantum case]
Let $\nu_1,\,\ldots,\,\nu_\frac{n}{2}$ be the symplectic eigenvalues of $\alpha$.
We can choose a symplectic basis of $\mathbb{R}^n$ in which
\begin{equation}
\alpha = \bigoplus_{i=1}^\frac{n}{2} \nu_i\,I_2\,.
\end{equation}
Let $X'$ be a quantum Gaussian system with $n/2$ modes, and for any $\nu\ge1/2$, let $\tau_{XX'}(\nu)$ be a quantum Gaussian state of $XX'$ such that $\tau_X(\nu) = \omega_X(\nu\,I_n)$.
We have
\begin{align}
S(X|M)\left(\mathcal{N}^X_{t\alpha}(\rho_{XM})\right) &\overset{\mathrm{(a)}}{\ge} \lim_{\nu\to\infty}S(X|X')\left(\mathcal{N}^X_{t\alpha}(\tau_{XX'}(\nu))\right) \overset{\mathrm{(b)}}{=} \sum_{i=1}^\frac{n}{2}\ln\left(\mathrm{e}\,t\,\nu_i\right)\nonumber\\
&\overset{\mathrm{(c)}}{=} \frac{1}{2}\ln\det\left(\mathrm{e}\,t\,\alpha\right)\,,
\end{align}
where (a) follows from [\onlinecite[Theorem 4]{de2018conditional}], (b) from [\onlinecite[Lemma 7]{de2018conditional}], and in (c) we have used that
\begin{equation}
\prod_{i=1}^\frac{n}{2}\nu_i = \sqrt{\det\alpha}\,.
\end{equation}
The claim follows.

\item[Classical case]
The conditional entropy is a concave function of the state [\onlinecite[Theorem 5.4]{ohya2004quantum}], therefore $S(X|M)\left(\mathcal{N}^X_{t\alpha}(\rho_{XM})\right)$ is minimized when $\rho_{XM}$ is pure, \emph{i.e.}, when $X$ has a deterministic value $x_0\in\mathbb{R}^n$ and $\rho_{XM} = \delta_X(x_0)\otimes\rho_M$, where $\delta_X(x_0)$ is a Dirac delta centered at $x_0$ and $\rho_M$ is pure.
We have
\begin{equation}
S(X|M)\left(\mathcal{N}^X_{t\alpha}(\delta_X(x_0)\otimes\rho_M)\right) = S_G(t\,\alpha) = \frac{1}{2}\ln\det\left(\mathrm{e}\,t\,\alpha\right)\,,
\end{equation}
and the claim follows.
\end{description}
\end{proof}

\section{Proof of Theorem 1}\label{sec:proof}
In this section we put together the Stam inequality proved in \autoref{sec:Stam} and the asymptotic scaling of the entropy proved in \autoref{sec:scaling} to prove our main result.

Let $(\mathbf{B},\mathbf{p})$ be a quantum Brascamp--Lieb datum.
For any $\rho_{XM}\in\mathcal{S}(XM)$, let
\begin{equation}
F^Q_{\mathbf{B},\mathbf{p}}(\rho_{XM}) = S(X|M)(\rho_{XM}) - \sum_{i=1}^K p_i\,S(Y_i|M)(\Phi_{B_i}(\rho_{XM}))\,,
\end{equation}
such that
\begin{equation}
f_Q(\mathbf{B},\mathbf{p}) = \sup_{\rho_{XM}\in\mathcal{S}(XM)}F^Q_{\mathbf{B},\mathbf{p}}(\rho_{XM})\,.
\end{equation}
First, we prove that if the scaling condition $2\,m = \mathbf{p}\cdot\mathbf{n}$ holds, the quantum Brascamp--Lieb constant is lower bounded by the classical Brascamp--Lieb constant:
\begin{prop}\label{prop:fgef*}
Let $\mathbf{B}$ be a quantum Brascamp--Lieb datum with dimension $(2m,\mathbf{n})$, and let $\mathbf{p}\in\mathbb{R}_{\ge0}^K$ satisfy the scaling condition $2\,m = \mathbf{p}\cdot\mathbf{n}$.
Then,
\begin{equation}
f_Q(\mathbf{B},\mathbf{p}) \ge f(\mathbf{B},\mathbf{p})\,.
\end{equation}
\end{prop} \begin{proof}
From \autoref{lem:SGas} of \autoref{app:SG}, for any $\alpha\in\mathbb{R}^{2m\times2m}_{>0}$ we have for $t\to\infty$
\begin{align}
S(X)(\omega_X(t\,\alpha)) &= m\ln t + \frac{1}{2}\ln\det\left(\mathrm{e}\,\alpha\right) + O\left(\frac{1}{t}\right)\,,\nonumber\\
S(Y_i)(\Phi_{B_i}(\omega_X(t\,\alpha))) &= S(Y_i)\left(\omega_{Y_i}\left(t\,B_i\,\alpha\,B_i^T\right)\right)\nonumber\\
&= m_i\ln t + \frac{1}{2}\ln\det\left(\mathrm{e}\,B_i\,\alpha\,B_i^T\right) + O\left(\frac{1}{t}\right)\,,
\end{align}
therefore
\begin{equation}\label{eq:fQlb}
f_Q(\mathbf{B},\mathbf{p}) \ge \lim_{t\to\infty}F^Q_{\mathbf{B},\mathbf{p}}(\omega_X(t\,\alpha)) = F_{\mathbf{B},\mathbf{p}}(\alpha)\,.
\end{equation}
The claim follows by taking the supremum of \eqref{eq:fQlb} over $\alpha\in\mathbb{R}^{2m\times2m}_{>0}$.
\end{proof}

We have to prove the converse inequality $f_Q(\mathbf{B},\mathbf{p}) \le f(\mathbf{B},\mathbf{p})$.
We can assume that $f(\mathbf{B},\mathbf{p})$ is finite.

\subsection{Achieved supremum in (\ref{eq:fdet})}
First, let us assume that $(\mathbf{B},\mathbf{p})$ satisfies any of the conditions of \autoref{thm:class}.
Let us fix $\rho_{XM}\in\mathcal{S}(XM)$, and let $\alpha\in\mathbb{R}^{2m\times2m}_{>0}$ achieve the supremum in \eqref{eq:fdet} and satisfy \eqref{eq:condalpha}.
For any $t\ge0$ and any $i=1,\,\ldots,\,K$, let
\begin{align}
\phi_X(t) &= S(X|M)\left(\mathcal{N}^X_{t\alpha}(\rho_{XM})\right)\,,\nonumber\\
\phi_{Y_i}(t) &= S(Y_i|M)\left(\Phi_{B_i}\left(\mathcal{N}^X_{t\alpha}(\rho_{XM})\right)\right) = S(Y_i|M)\left(\mathcal{N}^{Y_i}_{tB_i\alpha B_i^T}(\Phi_{B_i}(\rho_{XM}))\right)\,,
\end{align}
where the last equality follows from \autoref{lem:NPhi}, and let
\begin{equation}
\phi(t) = \phi_X(t) - \sum_{i=1}^K p_i\,\phi_{Y_i}(t) = F^Q_{\mathbf{B},\mathbf{p}}\left(\mathcal{N}^X_{t\alpha}(\rho_{XM})\right)\,.
\end{equation}
From \autoref{prop:as}, we have for $t\to\infty$
\begin{equation}
\phi_X(t) = \frac{1}{2}\ln\det\left(\mathrm{e}\,t\,\alpha\right) + O\left(\frac{1}{t}\right)\,,\qquad \phi_{Y_i}(t) = \frac{1}{2}\ln\det\left(\mathrm{e}\,t\,B_i\,\alpha\,B_i^T\right) + O\left(\frac{1}{t}\right)\,,
\end{equation}
and
\begin{equation}\label{eq:limt}
\lim_{t\to\infty}\phi(t) = \lim_{t\to\infty}\left(\frac{2\,m - \mathbf{p}\cdot\mathbf{n}}{2}\ln t + F_{\mathbf{B},\mathbf{p}}(\alpha) + O\left(\frac{1}{t}\right)\right) = F_{\mathbf{B},\mathbf{p}}(\alpha)\,,
\end{equation}
where we have used the scaling condition.
For any $i=1,\,\ldots,\,K$, let
\begin{equation}\label{eq:Pi}
\Pi_i = \alpha\,B_i^T\left(B_i\,\alpha\,B_i^T\right)^{-1}B_i\,.
\end{equation}
We have
\begin{equation}\label{eq:Pi2}
\Pi_i^2 = \Pi_i\,,\qquad \Pi_i\,\alpha = \alpha\,\Pi_i^T\,,\qquad B_i\,\Pi_i = B_i\,,
\end{equation}
and we get from \eqref{eq:condalpha}
\begin{equation}
\sum_{i=1}^K p_i\,\Pi_i = I_{2m}\,,
\end{equation}
such that
\begin{equation}\label{eq:Pialpha}
\sum_{i=1}^K p_i\,\Pi_i\,\alpha\,\Pi_i^T = \sum_{i=1}^K p_i\,\Pi_i^2\,\alpha = \sum_{i=1}^K p_i\,\Pi_i\,\alpha = \alpha\,.
\end{equation}
From \autoref{prop:tconcave}, $\phi$ is a linear combination of continuous and concave functions, and therefore it is continuous and almost everywhere differentiable.
We have for any $t\ge0$
\begin{align}\label{eq:ineqdB}
\phi_{Y_i}'(t) &\overset{\mathrm{(a)}}{=} J_{Y_i|M}\left(\mathcal{N}^{Y_i}_{tB_i\alpha B_i^T}(\Phi_{B_i}(\rho_{XM}))\right)\left(B_i\,\alpha\,B_i^T\right)\nonumber\\
&\overset{\mathrm{(b)}}{=} J_{Y_i|M}\left(\Phi_{B_i}\left(\mathcal{N}^X_{t\alpha}(\rho_{XM})\right)\right)\left(B_i\,\alpha\,B_i^T\right)\nonumber\\
&\overset{\mathrm{(c)}}{=} J_{Y_i|M}\left(\Phi_{B_i}\left(\mathcal{N}^X_{t\alpha}(\rho_{XM})\right)\right)\left(B_i\,\Pi_i\,\alpha\,\Pi_i^T\,B_i^T\right) \overset{\mathrm{(d)}}{\le} J_{X|M}\left(\mathcal{N}^X_{t\alpha}(\rho_{XM})\right)\left(\Pi_i\,\alpha\,\Pi_i^T\right)\,,
\end{align}
where (a) follows from the de Bruijn identity (\autoref{prop:dBJ}), (b) from \autoref{lem:NPhi}, (c) from \eqref{eq:Pi2} and (d) from the Stam inequality (\autoref{prop:Stam}).
We then have
\begin{align}
\phi'(t) &= \phi_X'(t) - \sum_{i=1}^n \phi_{Y_i}'(t) \ge \phi_X'(t) - \sum_{i=1}^K p_i\,J_{X|M}\left(\mathcal{N}^X_{t\alpha}(\rho_{XM})\right)\left(\Pi_i\,\alpha\,\Pi_i^T\right)\nonumber\\
&\overset{\mathrm{(a)}}{=} \phi_X'(t) - J_{X|M}\left(\mathcal{N}^X_{t\alpha}(\rho_{XM})\right)\left(\sum_{i=1}^K p_i\,\Pi_i\,\alpha\,\Pi_i^T\right)\nonumber\\
&\overset{\mathrm{(b)}}{=} \phi_X'(t) - J_{X|M}\left(\mathcal{N}^X_{t\alpha}(\rho_{XM})\right)(\alpha) \overset{\mathrm{(c)}}{=} 0\,,
\end{align}
where (a) follows from the linearity of the Fisher information (\autoref{prop:linearJ}), (b) from \eqref{eq:Pialpha} and (c) from the de Bruijn identity (\autoref{prop:dBJ}).
Therefore, $\phi$ is increasing, and \eqref{eq:limt} implies
\begin{equation}
F^Q_{\mathbf{B},\mathbf{p}}(\rho_{XM}) = \phi(0) \le \lim_{t\to\infty}\phi(t) = F_{\mathbf{B},\mathbf{p}}(\alpha) = f(\mathbf{B},\mathbf{p})\,.
\end{equation}
Since $\rho_{XM}$ is arbitrary, we conclude that
\begin{equation}\label{eq:fQlef}
f_Q(\mathbf{B},\mathbf{p}) \le f(\mathbf{B},\mathbf{p})\,.
\end{equation}
The claim follows.

\subsection{Generic case}
Let us now prove the inequality \eqref{eq:fQlef} in the generic case.
Since we have assumed that $f(\mathbf{B},\mathbf{p})$ is finite, by \autoref{thm:finite-class} any subspace of $\mathbb{R}^{2m}$ is subcritical for the pair $(\mathbf{B},\mathbf{p})$.

Let us fix an auxiliary pair $(\mathbf{B}',\mathbf{p}')$ whose only critical subspaces are $\mathbb{R}^{2m}$ and $\{0\}$.
For example, denoting with $e_1,\,\ldots,\,e_{2m}$ the row vectors of the canonical basis of $\mathbb{R}^{2m}$, we can choose
\begin{equation}
B_i' = e_i\quad \mathrm{for}\quad i=1,\,\ldots,\,2m\,,\qquad B_{2m+1}' = \sum_{i=1}^{2m} e_i
\end{equation}
and
\begin{equation}
p_1' = \ldots = p_{2m+1}' = \frac{2m}{2m+1}\,.
\end{equation}
We claim that for any subspace $\mathcal{V}\subset\mathbb{R}^{2m}$ with $\dim\mathcal{V} =k$, $0<k <2m$, \eqref{eq:Vsc} holds with strict inequality. Indeed, for every $i=1, \, \ldots,\, 2m+1$, $\dim B_i\mathcal{V} \in \left\{ 0,1\right\}$ and $\dim B_i \mathcal{V} = 1$ at least for $k+1$ different $i$'s, otherwise $\mathcal{V}$ would be orthogonal to the subspace generated by $2m+1-k$ vectors among $e_1, \, \ldots, \, e_{2m}, \sum_{i=1}^{2m} e_i$, that are necessarily independent, hence
\begin{equation}
k = \dim \mathcal{V} \le 2m - \left(2m+1-k\right) = k-1\, ,
\end{equation}
 a contradiction.
 Since $k<2m$, we have
 \begin{equation}
 k< \left(k+1\right) \frac{2m}{2m+1}\,,
 \end{equation}
 therefore \eqref{eq:Vsc} holds with strict inequality as claimed.

For any $0 \le \varepsilon\le 1$, we consider the pair $\left(\tilde{\mathbf{B}},\mathbf{p}_\varepsilon\right)$ with Brascamp--Lieb datum $\tilde{\mathbf{B}} = \mathbf{B} \cup \mathbf{B}'$ and $\mathbf{p}_\varepsilon = \left( \left(1-\varepsilon\right) \mathbf{p},\, \varepsilon \, \mathbf{p}'\right)$, such that
\begin{equation}
f_Q \left(\mathbf{B},\mathbf{p}\right) = f_Q \left(\tilde{\mathbf{B}},\mathbf{p}_0\right)\,, \quad  f \left( \mathbf{B},\mathbf{p}\right) = f\left( \tilde{\mathbf{B}},\mathbf{p}_0\right)\,.
\end{equation}
For any $0<\varepsilon\le 1$ and any subspace $\mathcal{V} \subseteq \mathbb{R}^{2m}$ we have
\begin{equation}
\dim \mathcal{V} \le \sum_{i=1}^K p_i \dim B_i \mathcal{V}\,, \qquad\dim \mathcal{V} \le  \sum_{j=1}^{2m+1} p_j' \dim B_j' \mathcal{V}\, ,
\end{equation}
where the second inequality is strict unless $\mathcal{V} = \mathbb{R}^{2m}$ or $\mathcal{V} = \{0\}$.
It follows that
\begin{equation}
 \dim \mathcal{V}  = \left(1-\varepsilon\right) \dim \mathcal{V} + \varepsilon \dim \mathcal{V} \le \sum_{i=1}^K \left(1-\varepsilon\right) p_i \dim B_i \mathcal{V}+ \sum_{j=1}^{2m+1} \varepsilon\,p'_j \dim B_j' \mathcal{V}\, ,
\end{equation}
with strict inequalities or equalities in the same cases. It follows that Condition \ref{critical_complementary} of \autoref{thm:class} is satisfied, hence $f_Q\left(\tilde{\mathbf{B}},\mathbf{p}_\varepsilon\right) = f\left(\tilde{\mathbf{B}},\mathbf{p}_\varepsilon\right)$.
Before taking the limit $\varepsilon\to0$, we need to prove the following continuity properties:
\begin{prop}\label{prop:sc}
For any Brascamp--Lieb datum $\mathbf{B}$, the function $\mathbf{p}\mapsto f(\mathbf{B},\mathbf{p})$ is convex and lower semicontinuous in $\mathbb{R}_{\ge0}^K$.
Moreover, let $\mathcal{R}\subseteq\mathbb{R}_{\ge0}^K$ be the intersection of a finite number of closed half-spaces such that $f(\mathbf{B},\mathbf{p})$ is finite for any $\mathbf{p}\in\mathcal{R}$.
Then, the function $\mathbf{p}\mapsto f(\mathbf{B},\mathbf{p})$ is continuous in $\mathcal{R}$.
If $\mathbf{B}$ is also a quantum Brascamp--Lieb datum, the function $\mathbf{p}\mapsto f_Q(\mathbf{B},\mathbf{p})$ enjoys the same properties.
\end{prop} \begin{proof}
Convexity and lower semicontinuity follow since the functions $\mathbf{p}\mapsto f(\mathbf{B},\mathbf{p})$ and $\mathbf{p}\mapsto f_Q(\mathbf{B},\mathbf{p})$ are suprema of affine functions. Continuity on  $\mathcal{R}$ is a consequence of [\onlinecite{gale1968convex}].
\end{proof}

\autoref{prop:sc} with $\mathcal{R} = \left\{ \mathbf{p}_\varepsilon\right\}_{\varepsilon \in [0,1]}$ yields
\begin{align}
f_Q(\mathbf{B},\mathbf{p}) &= f_Q \left( \tilde{\mathbf{B}},\mathbf{p}_0\right) \le \liminf_{\varepsilon \to 0} f_Q \left( \tilde{\mathbf{B}},\mathbf{p}_\varepsilon\right) = \lim_{\varepsilon \to 0} f \left( \tilde{\mathbf{B}},\mathbf{p}_\varepsilon\right) = f\left(\tilde{\mathbf{B}},\mathbf{p}_0\right)\nonumber\\
&= f(\mathbf{B},\mathbf{p})\,.
\end{align}
The claim follows.

\section{Applications}\label{sec:appl}

\subsection{Entropic uncertainty relations}\label{sec:EUR}
Entropic uncertainty relations provide a lower bound to the sum of the entropies of the outcomes of the measurements of non-commuting observables and play a key role in quantum information theory [\onlinecite{coles2017entropic}].
Their main applications are the proofs of the security of the protocols for quantum key distribution [\onlinecite{scarani2009security,pirandola2020advances}].
The main tool in this regard is provided by the uncertainty relations with quantum memory [\onlinecite{berta2010uncertainty}], where the entropies of the outcomes of the measurements are conditioned on a memory quantum system which plays the role of the adversary which wants to get information on the key.
The most common choice of non-commuting observables for the quantum Gaussian system $X$ with $m$ modes are the position $Q_X$ and the momentum $P_X$, which are given by the odd and even quadratures of $X$, respectively:
\begin{equation}
Q_i^X = R_{2i-1}^X,\,\qquad P_i^X = R_{2i}^X\,,\qquad i=1,\,\ldots,\,m\,.
\end{equation}
A fundamental entropic uncertainty relations with quantum memory for position and momentum proved by Frank and Lieb [\onlinecite{frank2013extended}] states that any $\rho_{XM}\in\mathcal{S}(SM)$ satisfies
\begin{equation}\label{eq:EUR}
S(X|M) \le S(Q_X|M) + S(P_X|M)\,.
\end{equation}
The inequality \eqref{eq:EUR} is a special case of the inequality \eqref{eq:QSSA} with $K=2$, $p_1 = p_2 = 1$ and where $Y_1$ and $Y_2$ are the random variables associated to the outcome of the measurement of position and momentum, respectively.
For a quantum Brascamp--Lieb datum $\mathbf{B}$ where each $B_i$ satisfies \eqref{eq:Bcomm}, each $Y_i$ is the random variable associated to the outcome of the measurement of the quadratures associated to the corresponding $B_i$.
Then, the generalized strong subadditivity \eqref{eq:QSSA} provides a lower bound to a linear combination of the conditional entropies of such random variables and therefore constitutes an entropic uncertainty relation with quantum memory which generalizes \eqref{eq:EUR}.

In the special case of a quantum Brascamp--Lieb datum $\mathbf{B}$ where each $B_i$ has rank one (\emph{i.e.}, it is a row vector in $\mathbb{R}^{2m}$), the Brascamp--Lieb constant $f(\mathbf{B}, \mathbf{p})$ is finite if and only if $\mathbf{p} \in \mathbb{R}_{\ge 0}^K$ is a convex combination of vectors of the form
\begin{equation}
p^{(\mathcal{I})}_i = \left\{
                        \begin{array}{cc}
                          1 & \quad i\in\mathcal{I} \\
                          0 & \quad i\notin\mathcal{I} \\
                        \end{array}
                      \right.\,,
\end{equation}
where $\mathcal{I}$ is any subset of $\left\{1, \ldots, K\right\}$ such that the row vectors $\left\{B_i:i \in \mathcal{I}\right\}$ form a basis of $\mathbb{R}^{2m}$ [\onlinecite[Theorem 5.5]{bennett2008brascamp}].
By \autoref{thm:main}, the same holds for the quantum Brascamp--Lieb constant $f_Q(\mathbf{B}, \mathbf{p})$, provided that the scaling condition $\sum_{i=1}^K p_i = 2m$ holds.

\subsection{Generalized quantum Entropy Power Inequality}\label{sec:EPI}
Let $X = X_1X_2$, where $X_1$ and $X_2$ are quantum Gaussian systems with $m$ modes each.
Given $\theta\in\mathbb{R}$, let $Y$ be the quantum Gaussian system with $m$ modes identified by the quadratures
\begin{equation}\label{eq:RYEPI}
R^Y_i = \cos\theta\,R^{X_1}_i + \sin\theta\,R^{X_2}_i\,,\qquad i=1,\,\ldots,\,2m\,.
\end{equation}
The \emph{quantum conditional Entropy Power Inequality} [\onlinecite{de2018conditional}] states that, for any $\rho_{X_1X_2M}\in\mathcal{S}(X_1X_2M)$ such that $X_1$ and $X_2$ are conditionally independent given $M$, \emph{i.e.},
\begin{equation}\label{eq:CI}
I(X_1;X_2|M)=0\,,
\end{equation}
we have
\begin{equation}\label{eq:EPI}
\mathrm{e}^\frac{S(Y|M)}{m} \ge \left(\cos\theta\right)^2\mathrm{e}^\frac{S(X_1|M)}{m} + \left(\sin\theta\right)^2\mathrm{e}^\frac{S(X_2|M)}{m}\,.
\end{equation}

In \autoref{prop:GEPI}, we will show that the generalized strong subadditivity of the von Neumann entropy \eqref{eq:GSSA} implies a generalization of the Entropy Power Inequality \eqref{eq:EPI} which does not require the conditional independence condition \eqref{eq:CI}.
We consider the following quantum Brascamp--Lieb datum:
\begin{equation}\label{eq:BLD}
B_1 = \left(
          \begin{array}{cc}
            I_{2m} & 0_{2m\times2m} \\
          \end{array}
        \right)\,,\qquad
B_2 = \left(
          \begin{array}{cc}
            0_{2m\times2m} & I_{2m} \\
          \end{array}
        \right)\,,\qquad
B_3 = \left(
          \begin{array}{cc}
            A_1 & A_2 \\
          \end{array}
        \right)\,,
\end{equation}
where $A_1,\,A_2\in\mathbb{R}^{2m\times2m}$ satisfy
\begin{equation}
A_1\,\Delta_{2m}\,A_1^T + A_2\,\Delta_{2m}\,A_2^T = \Delta_{2m}\,.
\end{equation}
$B_1$ and $B_2$ identify the subsystems $X_1$ and $X_2$, respectively, while $B_3$ identifies the subsystem $Y$ with $m$ modes associated to the following linear combination of the quadratures of $X$:
\begin{equation}
R^Y_i = \sum_{j=1}^{2m}\left(A_1\right)_{ij}R^{X_1}_i + \sum_{j=1}^{2m}\left(A_2\right)_{ij}R^{X_2}_i\,,\qquad i=1,\,\ldots,\,2m\,.
\end{equation}
\begin{rem}
For
\begin{equation}
A_1 = \cos\theta\,I_{2m}\,,\qquad A_2 = \sin\theta\,I_{2m}\,,
\end{equation}
$Y$ is the subsystem associated to the quadratures \eqref{eq:RYEPI}.
\end{rem}

The main idea to get such generalization of the Entropy Power Inequality \eqref{eq:EPI} is to optimize over $\mathbf{p}$ the generalized strong subadditivity \eqref{eq:QSSA}.
We need to define such optimization for all the classical Brascamp--Lieb data:
\begin{defn}
Let $\mathbf{B}$ be a Brascamp--Lieb datum with dimension $(n,\mathbf{n})$.
We define for any $\mathbf{s}\in\mathbb{R}^K$
\begin{equation}\label{eq:defphi}
\phi_{\mathbf{B}}(\mathbf{s}) = \inf\left(\mathbf{p}\cdot\mathbf{s} + f(\mathbf{B},\mathbf{p}):\mathbf{p}\in\mathbb{R}_{\ge0}^K\,,\; n = \mathbf{p}\cdot\mathbf{n}\right)\,.
\end{equation}
\end{defn}

Then, the following inequality holds (a similar inequality holds in the classical case):
\begin{prop}\label{prop:phiQ}
Let $\mathbf{B}$ be a quantum Brascamp--Lieb datum with dimension $(2m,\mathbf{n})$ as in \autoref{defn:QBL}.
Then, for any $\rho_{XM}\in\mathcal{S}(XM)$ we have
\begin{equation}
S(X|M)(\rho_{XM}) \le \phi_{\mathbf{B}}\left(S(Y_1|M)(\Phi_{B_1}(\rho_{XM})),\,\ldots,\,S(Y_K|M)(\Phi_{B_K}(\rho_{XM}))\right)\,.
\end{equation}
\end{prop} \begin{proof}
We have for any $\mathbf{p}\in\mathbb{R}_{\ge0}^K$ such that $2\,m = \mathbf{p}\cdot\mathbf{n}$
\begin{align}
S(X|M)(\rho_{XM}) &\le \sum_{i=1}^K p_i\,S(Y_i|M)(\Phi_{B_i}(\rho_{XM})) + f_Q(\mathbf{B},\mathbf{p})\nonumber\\
&= \sum_{i=1}^K p_i\,S(Y_i|M)(\Phi_{B_i}(\rho_{XM})) + f(\mathbf{B},\mathbf{p})\,,
\end{align}
where the inequality follows from the definition of $f_Q$, and the equality from \autoref{thm:main}.
The claim follows taking the infimum over $\mathbf{p}$.
\end{proof}

In the following, we will determine the function $\phi_\mathbf{B}$ for the quantum Brascamp--Lieb datum \eqref{eq:BLD}.

If $X$ is a random variable with values in $\mathbb{R}^n$ and $A$ is an invertible $n\times n$ real matrix, the entropy of $AX$ is related to the entropy of $X$ by
\begin{equation}
S(AX) = S(X) + \ln\left|\det A\right|\,.
\end{equation}
This property leads us to introduce the notion of equivalent Brascamp--Lieb data:

\begin{defn}\label{defn:equiv}
The Brascamp--Lieb data $\mathbf{B}$ and $\mathbf{B}^0$ with dimension $(n,\mathbf{n})$ are \emph{equivalent} if for any $i=1,\,\ldots,\,K$ we have
\begin{equation}
B_i = A_i^{-1}\,B^0_i\,A
\end{equation}
for some $A\in \mathrm{GL}_n(\mathbb{R})$ and $A_i\in \mathrm{GL}_{n_i}(\mathbb{R})$.
\end{defn}

The Brascamp--Lieb constants of equivalent data are easily related:
\begin{prop}\label{prop:BB0}
Let $\mathbf{B}$ and $\mathbf{B}^0$ be equivalent Brascamp--Lieb data as in \autoref{defn:equiv}.
Then, for any $\mathbf{p}\in\mathbb{R}_{\ge0}^K$ and any $\mathbf{s}\in\mathbb{R}^K$ we have
\begin{align}
f(\mathbf{B},\mathbf{p}) &= f(\mathbf{B}^0,\mathbf{p}) - \ln\left|\det A\right| + \sum_{i=1}^K p_i\ln\left|\det A_i\right|\,,\nonumber\\
\phi_\mathbf{B}(\mathbf{s}) &= \phi_{\mathbf{B}^0}\left(s_1+\ln\left|\det A_1\right|,\,\ldots,\,s_K+\ln\left|\det A_K\right|\right) - \ln\left|\det A\right|\,.
\end{align}
\end{prop}

First, we compute the Brascamp--Lieb constant of a datum in a standard form, and then get from \autoref{prop:BB0} the constant of the equivalent data:

\begin{prop}\label{prop:B0}
We consider the following Brascamp--Lieb datum:
\begin{equation}\label{eq:B0}
B_1^0 = \left(
          \begin{array}{cc}
            I_n & 0_{n\times n} \\
          \end{array}
        \right)\,,\qquad
B_2^0 = \left(
          \begin{array}{cc}
            0_{n\times n} & I_n \\
          \end{array}
        \right)\,,\qquad
B_3^0 = \left(
          \begin{array}{cc}
            I_n & I_n \\
          \end{array}
        \right)\,.
\end{equation}
Then, for any $\mathbf{p}\in\mathbb{R}_{\ge0}^3$ we have
\begin{equation}
f(\mathbf{B}^0,\mathbf{p}) = \frac{n}{2}\sum_{i=1}^3\left(\left(1-p_i\right)\ln\left(1-p_i\right) - p_i\ln p_i\right)
\end{equation}
if $\mathbf{p}\in[0,1]^3$ and $p_1+p_2+p_3=2$, and $f(\mathbf{B}^0,\mathbf{p})=\infty$ otherwise.
\end{prop} \begin{proof}
The condition $p_1+p_2+p_3=2$ is the scaling condition.
The subcriticality of $\ker B_1^0$ implies
\begin{equation}
1 \le p_2 + p_3 = 2 - p_1\,,
\end{equation}
therefore $p_1\le1$.
Analogously, the subcriticality of $\ker B_2^0$ and $\ker B_3^0$ imply $p_2\le1$ and $p_3\le1$, respectively.
If $\mathbf{p}\in(0,1)^3$, condition \eqref{eq:condalpha} is satisfied by
\begin{equation}
\alpha = \left(
           \begin{array}{cc}
             p_1\left(1-p_1\right)I_n & -\left(1-p_1\right)\left(1-p_2\right)I_n \\
             -\left(1-p_1\right)\left(1-p_2\right)I_n & p_2\left(1-p_2\right)I_n \\
           \end{array}
         \right) > 0\,.
\end{equation}
Therefore, \autoref{thm:class} implies
\begin{equation}
f(\mathbf{B}^0,\mathbf{p}) = F_{\mathbf{B}^0,\mathbf{p}}(\alpha) = \frac{n}{2}\sum_{i=1}^3\left(\left(1-p_i\right)\ln\left(1-p_i\right) - p_i\ln p_i\right)\,.
\end{equation}
The extension to $[0,1]^3$ follows by continuity from \autoref{prop:sc}.
\end{proof}

Let us determine the function $\phi_{\mathbf{B}^0}$:
\begin{prop}\label{prop:phi0}
Let $\mathbf{B}^0$ be a Brascamp--Lieb datum as in \eqref{eq:B0}, let $\mathbf{s}\in\mathbb{R}^3$ and for $i=1,\,2,\,3$, let $e_i = \mathrm{e}^\frac{2s_i}{n}$.
If
\begin{equation}\label{eq:triangle}
e_1<e_2 + e_3\,,\qquad e_2<e_1+e_3\,,\qquad e_3 < e_1+e_2\,,
\end{equation}
then
\begin{align}
&\phi_{\mathbf{B}^0}(\mathbf{s}) =\nonumber\\
&\frac{n}{2}\ln\frac{\left(e_2 + e_3 - e_1\right)e_1 + \left(e_1 + e_3 - e_2\right)e_2 + \left(e_1 + e_2 - e_3\right)e_3}{4}\,;
\end{align}
otherwise,
\begin{equation}
\phi_{\mathbf{B}^0}(\mathbf{s}) = \min\left(s_1 + s_2,\,s_1 + s_3,\,s_2+s_3\right)\,.
\end{equation}
\end{prop} \begin{proof}
For \autoref{prop:B0}, we can restrict the supremum in \eqref{eq:defphi} to $\mathbf{p}\in[0,1]^3:p_1+p_2+p_3=2$.
For any $p_1,\,p_2\in[0,1]$ such that $p_1+p_2\ge1$, let
\begin{equation}
\psi(p_1,p_2) = \left.\left(\mathbf{s}\cdot\mathbf{p} + f(\mathbf{B}^0,\mathbf{p})\right)\right|_{p_3 = 2 - p_1 - p_2}\,.
\end{equation}
We have
\begin{align}
\frac{\partial\psi}{\partial p_1}(p_1,p_2) &= \frac{n}{2}\ln\frac{e_1\left(2-p_1-p_2\right)\left(p_1+p_2-1\right)}{e_3\,p_1\left(1-p_1\right)}\,,\nonumber\\
\frac{\partial\psi}{\partial p_2}(p_1,p_2) &= \frac{n}{2}\ln\frac{e_2\left(2-p_1-p_2\right)\left(p_1+p_2-1\right)}{e_3\,p_2\left(1-p_2\right)}\,.
\end{align}
We have $\nabla\psi(\bar{p}_1,\bar{p}_2)=0$ for
\begin{align}
\bar{p}_1 &= \frac{2\left(e_2 + e_3 - e_1\right)e_1}{\left(e_2 + e_3 - e_1\right)e_1 + \left(e_1 + e_3 - e_2\right)e_2 + \left(e_1 + e_2 - e_3\right)e_3}\,,\nonumber\\
\bar{p}_2 &= \frac{2\left(e_1 + e_3 - e_2\right)e_2}{\left(e_2 + e_3 - e_1\right)e_1 + \left(e_1 + e_3 - e_2\right)e_2 + \left(e_1 + e_2 - e_3\right)e_3}\,.
\end{align}
If condition \eqref{eq:triangle} holds, we have $\bar{p}_1,\,\bar{p}_2\in(0,1)$ and $\bar{p}_1 + \bar{p}_2 > 1$, and since $\psi$ is convex, it achieves in $(\bar{p}_1,\bar{p}_2)$ its minimum.
Therefore,
\begin{align}
&\phi_{\mathbf{B}^0}(\mathbf{s}) = \psi(\bar{p}_1,\bar{p}_2)\nonumber\\
&= \frac{n}{2}\ln\frac{\left(e_2 + e_3 - e_1\right)e_1 + \left(e_1 + e_3 - e_2\right)e_2 + \left(e_1 + e_2 - e_3\right)e_3}{4}\,.
\end{align}
If \eqref{eq:triangle} does not hold, $\nabla\psi(p_1,p_2)\neq0$ for any $p_1,p_2\in(0,1)^3$ such that $p_1+p_2>1$.
Therefore, the supremum in \eqref{eq:defphi} can be restricted to the region
\begin{equation}
\left\{\left(1,p,1-p\right),\;\left(p,1,1-p\right),\;\left(p,1-p,1\right):p\in[0,1]\right\}\,.
\end{equation}
On such region we have $f(\mathbf{B}^0,\mathbf{p})=0$.
Without loss of generality, we can assume that $s_1\le s_2 \le s_3$.
Since \eqref{eq:triangle} does not hold, we must have
\begin{equation}
e_3\ge e_1+e_2\,,
\end{equation}
and the supremum in \eqref{eq:defphi} is achieved in $\mathbf{p}=(1,1,0)$.
The claim follows.
\end{proof}

We can finally prove the generalization of the Entropy Power Inequality \eqref{eq:EPI}:
\begin{prop}\label{prop:GEPI}
Let $\mathbf{B}$ and $\mathbf{B}^0$ be quantum Brascamp--Lieb data as in \eqref{eq:BLD} and \eqref{eq:B0} with $n=2m$, respectively, and let
\begin{equation}
\lambda_i = \left|\det A_i\right|^\frac{1}{m}\,,\qquad i=1,\,2\,.
\end{equation}
Let $\rho_{X_1X_2M}\in\mathcal{S}(X_1X_2M)$.
If $\rho_{X_1X_2M}$ satisfies the following three inequalities:
\begin{align}
\lambda_1\,\mathrm{e}^\frac{S(X_1|M)}{m} &< \lambda_2\,\mathrm{e}^\frac{S(X_2|M)}{m} + \mathrm{e}^\frac{S(Y|M)}{m}\,,\nonumber\\
\lambda_2\,\mathrm{e}^\frac{S(X_2|M)}{m} &<\lambda_1\,\mathrm{e}^\frac{S(X_1|M)}{m}+\mathrm{e}^\frac{S(Y|M)}{m}\,,\nonumber\\
\mathrm{e}^\frac{S(Y|M)}{m} &< \lambda_1\,\mathrm{e}^\frac{S(X_1|M)}{m}+\lambda_2\,\mathrm{e}^\frac{S(X_2|M)}{m}\,,
\end{align}
then
\begin{align}
4\,\lambda_1\,\lambda_2\,\mathrm{e}^\frac{S(X_1X_2|M)}{m} &\le \left(\lambda_2\,\mathrm{e}^\frac{S(X_2|M)}{m} + \mathrm{e}^\frac{S(Y|M)}{m} - \lambda_1\,\mathrm{e}^\frac{S(X_1|M)}{m}\right)\lambda_1\,\mathrm{e}^\frac{S(X_1|M)}{m}\nonumber\\
& \phantom{\le} + \left(\lambda_1\,\mathrm{e}^\frac{S(X_1|M)}{m} + \mathrm{e}^\frac{S(Y|M)}{m} - \lambda_2\,\mathrm{e}^\frac{S(X_2|M)}{m}\right)\lambda_2\,\mathrm{e}^\frac{S(X_2|M)}{m}\nonumber\\
& \phantom{\le} + \left(\lambda_1\,\mathrm{e}^\frac{S(X_1|M)}{m} + \lambda_2\,\mathrm{e}^\frac{S(X_2|M)}{m} - \mathrm{e}^\frac{S(Y|M)}{m}\right)\mathrm{e}^\frac{S(Y|M)}{m}\,;
\end{align}
otherwise,
\begin{align}
S(X_1X_2|M) &\le \min\left(S(X_1|M) + S(X_2|M),\,S(X_1|M) + S(Y|M) -\ln\left|\det A_2\right|,\right.\nonumber\\
&\phantom{\le \min} \; \left.S(X_2|M)+S(Y|M)-\ln\left|\det A_1\right|\right)\,.
\end{align}
\end{prop} \begin{proof}
$\mathbf{B}$ and $\mathbf{B}^0$ are equivalent as in \autoref{defn:equiv} with
\begin{equation}
A_3 = I_{2m}\,,\qquad A = \mathrm{diag}(A_1,\,A_2)\,.
\end{equation}
Therefore, \autoref{prop:B0} gives for any $\mathbf{s}\in\mathbb{R}^3$
\begin{equation}
\phi_{\mathbf{B}}(\mathbf{s}) = \phi_{\mathbf{B}^0}\left(s_1+\ln\left|\det A_1\right|,\,s_2+\ln\left|\det A_2\right|,\,s_3\right) - \ln\left|\det A_1\right| - \ln\left|\det A_2\right|\,
\end{equation}
and the claim follows from \autoref{prop:phi0}.
\end{proof}

\subsection{Entanglement production}\label{sec:ent}
Let $X = X_1X_2$, where $X_1$ and $X_2$ are quantum Gaussian systems with $m_1$ and $m_2$ modes, respectively.
Let
\begin{equation}\label{eq:S}
S = \left(
      \begin{array}{cc}
        S_{11} & S_{12} \\
        S_{21} & S_{22} \\
      \end{array}
    \right)
\end{equation}
be a symplectic matrix, \emph{i.e.}, a $(2m_1+2m_2)\times(2m_1+2m_2)$ real matrix such that
\begin{equation}\label{eq:sympl}
S\,\Delta_{2m_1+2m_2}\,S^T = \Delta_{2m_1+2m_2}\,,
\end{equation}
and let $U_S$ be the Gaussian symplectic unitary operator acting on $L^2(\mathbb{R}^{m_1+m_2})$ that performs the linear redefinition of the quadratures of $X_1X_2$ induced by $S$:
\begin{align}\label{eq:X'}
R_i^{X_1'} &= U_S^\dag\,R_i^{X_1}\,U_S = \sum_{j=1}^{2m_1} \left(S_{11}\right)_{ij}R^{X_1}_j + \sum_{j=1}^{2m_2} \left(S_{12}\right)_{ij}R^{X_2}_j\,,\qquad i=1,\,\ldots,\,2m_1\,,\nonumber\\
R_i^{X_2'} &= U_S^\dag\,R^{X_2}_i\,U_S = \sum_{j=1}^{2m_1} \left(S_{21}\right)_{ij}R^{X_1}_j + \sum_{j=1}^{2m_2} \left(S_{22}\right)_{ij}R^{X_2}_j\,,\qquad i=1,\,\ldots,\,2m_2\,.
\end{align}

The generalized strong subadditivity \eqref{eq:QSSA} provides a lower bound to the sum of the correlations between $X_1$ and $X_2$ before and after applying $U_S$:

\begin{prop}\label{prop:ent0}
Let $\mathbf{B}$ be the quantum Brascamp--Lieb datum
\begin{align}\label{eq:BS}
B_1 &= \left(
        \begin{array}{cc}
          I_{2m_1} & 0_{2m_1\times2m_2} \\
        \end{array}
      \right)\,,\qquad
B_2 = \left(
        \begin{array}{cc}
          0_{2m_2\times2m_1} & I_{2m_2} \\
        \end{array}
      \right)\,,\nonumber\\
B_1' &= \left(
        \begin{array}{cc}
          S_{11} & S_{12} \\
        \end{array}
      \right)\,,\qquad
B_2' = \left(
        \begin{array}{cc}
          S_{21} & S_{22} \\
        \end{array}
      \right)\,,
\end{align}
and let
\begin{equation}\label{eq:PS}
\mathbf{p} = \left(\frac{1}{2},\,\frac{1}{2},\,\frac{1}{2},\,\frac{1}{2}\right)\,.
\end{equation}
Then, for any quantum state $\rho_{X_1X_2M}\in\mathcal{S}(X_1X_2M)$ we have
\begin{equation}
\frac{1}{2}\,I(X_1;X_2|M)(\rho_{X_1X_2M}) + \frac{1}{2}\,I(X_1;X_2|M)\left(U_S\,\rho_{X_1X_2M}\,U_S^\dag\right) \ge -f(\mathbf{B},\mathbf{p})\,.
\end{equation}
\end{prop} \begin{proof}
$\mathbf{p}$ satisfies the scaling condition, therefore \autoref{thm:main} implies $f_Q(\mathbf{B},\mathbf{p}) = f(\mathbf{B},\mathbf{p})$.
$X_1X_2$ and $X_1'X_2'$ are related by the unitary transformation $U_S$, therefore
\begin{equation}
S(X_1X_2|M) = S(X_1'X_2'|M)\,,
\end{equation}
and
\begin{align}
&\frac{I(X_1;X_2|M) + I(X_1';X_2'|M)}{2}\nonumber\\
&= \frac{S(X_1|M) + S(X_2|M) - S(X_1X_2|M) + S(X_1'|M) + S(X_2'|M) - S(X_1'X_2'|M)}{2}\nonumber\\
&= \frac{S(X_1|M) + S(X_2|M) + S(X_1'|M) + S(X_2'|M)}{2} - S(X_1X_2|M) \ge -f_Q(\mathbf{B},\mathbf{p})\nonumber\\
&= -f(\mathbf{B},\mathbf{p})\,.
\end{align}
The claim follows.
\end{proof}

The \emph{entanglement entropy} of a pure state $\rho_{X_1X_2}$ of $X_1X_2$ is a measure of the quantum correlations between $X_1$ and $X_2$ given by $I(X_1;X_2)/2$.
\autoref{prop:ent0} applied to a pure state $\rho_{X_1X_2}$ with a trivial $M$ provides a lower bound to the sum of the entanglement entropies of $\rho_{X_1X_2}$ before and after applying $U_S$.
The entanglement entropy generated by a Gaussian symplectic unitary transformation has been studied in the case of Gaussian states when the transformation is generated by a time-independent quadratic Hamiltonian [\onlinecite{bianchi2015entanglement,bianchi2018linear,hackl2018entanglement}], \emph{i.e.}, where the symplectic matrix $S$ has the form
\begin{equation}\label{eq:St}
S(t) = \mathrm{e}^{tH}\,,\qquad t\in\mathbb{R}\,,
\end{equation}
where $H$ is a $(2m_1+2m_2)\times(2m_1+2m_2)$ real matrix satisfying\footnote{Condition \eqref{eq:HS} ensures that the matrix $S(t)$ is symplectic at all times.}
\begin{equation}\label{eq:HS}
H\,\Delta_{2m_1+2m_2} = \Delta_{2m_1+2m_2}\,H^T\,.
\end{equation}

Ref. [\onlinecite{hackl2018entanglement}] focuses on the limit $t\to\infty$ and proves that, for any Gaussian initial state, the entanglement entropy grows linearly with time.
The linear scaling of the entanglement entropy for generic initial states is left as an open conjecture, which is supported by numerical experiments.
In \autoref{prop:entS}, we prove such conjecture in the case where the matrix $H$ is symmetric.
First, we show that we can get an analytic expression for the Brascamp--Lieb constant if the symplectic matrix $S$ is positive definite:
\begin{prop}\label{prop:ent}
Let $S$ be a symplectic matrix as in \eqref{eq:S} which is also symmetric and positive definite.
Let $\mathbf{B}$ be the quantum Brascamp--Lieb datum \eqref{eq:BS}, and let $\mathbf{p}$ be as in \eqref{eq:PS}.
Then,
\begin{equation}
f(\mathbf{B},\mathbf{p}) = -\frac{1}{2}\ln\left(\det S_{11}\det S_{22}\right)\,.
\end{equation}
\end{prop} \begin{proof}
We have that $\alpha = S^{-1}$ satisfies \eqref{eq:condalpha}, therefore \autoref{thm:class} implies
\begin{equation}
f(\mathbf{B},\mathbf{p}) = F_{\mathbf{B},\mathbf{p}}\left(S^{-1}\right) = -\frac{1}{2}\ln\left(\det S_{11}\det S_{22}\right)\,.
\end{equation}
The claim follows.
\end{proof}

We can now prove the conjecture:
\begin{prop}\label{prop:entS}
Let $H$ and $S(t)$ be as in \eqref{eq:HS} and \eqref{eq:St}, respectively.
Let us suppose that $H$ is symmetric and has block decomposition
\begin{equation}
H = \left(
      \begin{array}{cc}
        H_{11} & H_{12} \\
        H_{21} & H_{22} \\
      \end{array}
    \right)
\end{equation}
with $H_{12} \neq 0 $.
Then, there exists $\Lambda>0$ depending only on $H$ and such that for any pure state $\rho_{X_1X_2}\in\mathcal{S}(X_1X_2)$ we have for $t\to\infty$
\begin{equation}
\frac{1}{2}\,I(X_1;X_2)\left(U_{S(t)}\,\rho_{X_1X_2}\,U_{S(t)}^\dag\right) \ge \Lambda\,t + O(1)\,.
\end{equation}
\end{prop} \begin{proof}
Since $H$ is symmetric, $S(t)$ is symmetric and positive definite for any $t\in\mathbb{R}$.
Then, \autoref{prop:ent0} and \autoref{prop:ent} imply
\begin{equation}
\frac{1}{2}\,I(X_1;X_2)\left(U_{S(t)}\,\rho_{X_1X_2}\,U_{S(t)}^\dag\right) \ge \frac{1}{2}\ln f(t) - \frac{1}{2}\,I(X_1;X_2)(\rho_{X_1X_2})\,,
\end{equation}
where for any $t\in\mathbb{R}$,
\begin{equation}
f(t) = \det S(t)_{11}\det S(t)_{22}\,.
\end{equation}
Since $H$ is symmetric, its eigenvalues $\lambda_1,\,\ldots,\,\lambda_{2m_1+2m_2}$ are real, and there exists an orthogonal matrix $V\in\mathrm{O}_{2m_1+2m_2}(\mathbb{R})$ such that
\begin{equation}
H = V\,\mathrm{diag}\left(\lambda_1,\,\ldots,\,\lambda_{2m_1+2m_2}\right)V^T\,.
\end{equation}
We have for any $t\in\mathbb{R}$
\begin{equation}
S(t) = V\,\mathrm{diag}\left(\mathrm{e}^{t\lambda_1},\,\ldots,\,\mathrm{e}^{t\lambda_{2m_1+2m_2}}\right)V^T\,,
\end{equation}
therefore, the entries of $S(t)$ are linear functions of $\mathrm{e}^{t\lambda_1},\,\ldots,\,\mathrm{e}^{t\lambda_{2m_1+2m_2}}$, and $f(t)$ is a homogeneous polynomial of degree $2m_1+2m_2$ in $\mathrm{e}^{t\lambda_1},\,\ldots,\,\mathrm{e}^{t\lambda_{2m_1+2m_2}}$.
Let $C\,\mathrm{e}^{t\Lambda}$ be the dominant monomial in $f(t)$ for $t\to\infty$, such that
\begin{equation}
\ln f(t) = \Lambda\,t + O(1)\,.
\end{equation}
The claim follows if we prove that $\Lambda>0$.
First, we claim that $f(-t) = f(t)$.
Indeed, since $S(t)$ is symmetric and satisfies \eqref{eq:sympl}, we have
\begin{equation}
S(-t) = {S(t)}^{-1} = \Delta_{2m_1+2m_2}\,S(t)\,\Delta_{2m_1+2m_2}^{-1}\,,
\end{equation}
therefore,
\begin{equation}
\det S(-t)_{11} = \det\left(\Delta_{2m_1}\,S(t)_{11}\,\Delta_{2m_1}^{-1}\right) = \det S(t)_{11}\,,
\end{equation}
and the same property holds for $S(t)_{22}$.
Then, if $f(t)$ contains the monomial $C\,\mathrm{e}^{t\Lambda}$, it contains the monomial $C\,\mathrm{e}^{-t\Lambda}$ as well, hence $\Lambda\ge0$.
Therefore, the claim follows if we prove that $\Lambda\neq0$.
Let us assume that $\Lambda=0$.
This can happen only if $f(t)$ is constant, \emph{i.e.}, if $f(t) = f(0)=1$ for any $t\in\mathbb{R}$.
Since $S(t)$ is positive definite, $S(t)_{11}$ and $S(t)_{22}$ are positive definite as well, and therefore invertible.
Any symplectic matrix has unit determinant, and from the formula for the determinant of a block matrix we get
\begin{align}\label{eq:det1}
0 &= \ln\det S(t) = \ln\left(\det S(t)_{11}\det S(t)_{22} \det\left(I_{2m_1} - {S(t)}_{11}^{-1}\,S(t)_{12}\,{S(t)}_{22}^{-1}\,S(t)_{21}\right)\right)\nonumber\\
&= \ln\det\left(I_{2m_1} - {S(t)}_{11}^{-1}\,S(t)_{12}\,{S(t)}_{22}^{-1}\,S(t)_{21}\right) \le -\ \mathrm{tr}\left[{S(t)}_{11}^{-1}\,S(t)_{12}\,{S(t)}_{22}^{-1}\,S(t)_{21}\right]\nonumber\\
&= -\left\|{S(t)}_{11}^{-\frac{1}{2}}\,S(t)_{12}\,{S(t)}_{22}^{-\frac{1}{2}}\right\|_2^2\,,
\end{align}
where
\begin{equation}
\left\|A\right\|_2 = \sqrt{\mathrm{tr}\left[A\,A^T\right]}
\end{equation}
denotes the Frobenius norm.
The identity \eqref{eq:det1} can hold iff
\begin{equation}
{S(t)}_{11}^{-\frac{1}{2}}\,S(t)_{12}\,{S(t)}_{22}^{-\frac{1}{2}} = 0
\end{equation}
for any $t\in\mathbb{R}$, which implies
\begin{equation}
H_{12} = S'(0)_{12} = 0\,,
\end{equation}
contradicting the hypothesis $H_{12}\neq0$.
The claim follows.
\end{proof}

\begin{acknowledgments}
G.D.P. has been supported by the HPC Italian National Centre for HPC, Big Data and Quantum Computing - Proposal code CN00000013 - CUP J33C22001170001 under the MUR National Recovery and Resilience Plan funded by the European Union - NextGenerationEU.
G.D.P. is a member of the ``Gruppo Nazionale per la Fisica Matematica (GNFM)'' of the ``Istituto Nazionale di Alta Matematica ``Francesco Severi'' (INdAM)''.
D.T. was partially supported by INdAM Gnampa project 2019 ``Problemi di ottimizzazione con vincoli via trasporto ottimo e incertezza''.
\end{acknowledgments}

\appendix

\section{Classical Brascamp--Lieb inequalities}\label{app:BL}

Let $\mathbf{B}$ be  a Brascamp--Lieb datum with dimension $(n, \mathbf{n}) \in \mathbb{N} \times\mathbb{N}^K$ and let $\mathbf{p} \in \mathbb{R}^K_{\ge 0}$. The classical Brascamp--Lieb inequality [\onlinecite{bennett2008brascamp}] states that
\begin{equation}\label{eq:classical-BL} \int_{\mathbb{R}^n} \prod_{i=1}^K {f_i( B_i x )}^{p_i}\,\mathrm{d}^nx \le \operatorname{BL}(\mathbf{B}, \mathbf{p}) \prod_{i=1}^K \left( \int_{\mathbb{R}^{n_i}} f_i(y_i) \,\mathrm{d}^{n_i}y_i\right)^{p_i}\,,
\end{equation}
where $0 \le \operatorname{BL}(\mathbf{B}, \mathbf{p}) \le \infty$ is defined as the smallest constant such that the inequality holds for all measurable functions $f_i: \mathbb{R}^{n_i} \to [0, \infty)$. The inequality \eqref{eq:classical-BL} was first proved in [\onlinecite{brascamp1976best}] in the case $n_i= 1$ and generalizes several functional inequalities, such as the multilinear H\"older inequality, the sharp Young convolution inequality [\onlinecite{brascamp1976best,barthe1998optimal}] and the Loomis–Whitney inequality [\onlinecite{loomis1949inequality}].
The Brascamp--Lieb inequality has also been generalized to groups and homogeneous spaces [\onlinecite{barthe2011correlation}].

The link between the Brascamp--Lieb inequality \eqref{eq:classical-BL} and the generalized strong subadditivity of the Shannon differential entropy \eqref{eq:GSSA} was first noticed in [\onlinecite{carlen2004sharp}] and later developed in [\onlinecite{carlen2009subadditivity}], where it was proved that [\onlinecite[Theorem 2.1]{carlen2009subadditivity}]
\begin{equation}
\operatorname{BL}(\mathbf{B}, \mathbf{p}) = \exp\,f(\mathbf{B}, \mathbf{p})\, ,
\end{equation}
together with a precise correspondence between optimizers, whenever they exist [\onlinecite[Theorem 2.2]{carlen2009subadditivity}]. Several equivalent conditions for finiteness and existence of optimizers are given in [\onlinecite{bennett2008brascamp}]. See also [\onlinecite{liu2016brascamp,liu2017information,courtade2021euclidean}] for recent developments.

Refs. [\onlinecite{carlen2008brascamp,berta2019quantum}] propose quantum generalizations of \eqref{eq:classical-BL}, where nonnegative functions are replaced by positive semidefinite linear operators and integrals are replaced by traces.
Refs. [\onlinecite{carlen2008brascamp,berta2019quantum}] also prove that each quantum Brascamp--Lieb inequality is equivalent to a corresponding generalization of the subadditivity of the von Neumann entropy.
Such equivalence also applies to the generalized subadditivity of the von Neumann entropy proved in this paper:
[\onlinecite[Corollary II.5]{berta2019quantum}] and \autoref{thm:main} yield the following
\begin{cor}
Let $\mathbf{B}$ be a quantum Brascamp--Lieb datum with dimension $(2m,2\mathbf{m})$ as in \autoref{defn:QBL} such that each $B_i$ satisfies \eqref{eq:DeltaY}, and let $\mathbf{p}\in\mathbb{R}_{>0}^K$ satisfy the scaling condition $m=\mathbf{p}\cdot \mathbf{m}$.
For any $i=1,\,\ldots,\,K$, let $\omega_i$ be a positive definite linear operator acting on $L^2(\mathbb{R}^{m_i})$ such that $\ln \omega_i$ is in the domain of $\Phi_{B_i}^\dagger$\footnote{For any quantum channel $\Phi$ that maps quantum states acting on the Hilbert space $\mathcal{H}_X$ to quantum states acting on the Hilbert space $\mathcal{H}_Y$, we denote with $\Phi^\dag$ the dual quantum channel that maps bounded linear operators acting on $\mathcal{H}_Y$ to bounded linear operators acting on $\mathcal{H}_X$ and such that $\mathrm{Tr}\left[\Phi^\dag(H_Y)\,\rho_X\right] = \mathrm{Tr}\left[H_Y\,\Phi(\rho_X)\right]$ for any quantum state $\rho_X$ acting on $\mathcal{H}_X$ and any bounded linear operator $H_Y$ acting on $\mathcal{H}_Y$.}.
Then,
\begin{equation}
\mathrm{Tr}\left[ \exp \left( \sum_{i=1}^K \Phi_{B_i}^\dagger( \ln \omega_i )  \right) \right ] \le \exp f(\mathbf{B}, \mathbf{p})\; \prod_{i=1}^K \left(\mathrm{Tr}\,\omega_i^{\frac{1}{p_i}} \right)^{p_i}\, .
\end{equation}
\end{cor}

\section{Entropy of the joint measurement of commuting quadratures}\label{app:entropy}

\begin{lem}\label{lem:finite-energy-classical}
Let $\rho_X$ be a quantum state with finite energy of a bosonic quantum Gaussian system with $m$ modes, let $B$ be an $n \times 2m$ matrix with rank $n$ satisfying \eqref{eq:Bcomm} and let $Y$ denote the random variable obtained by jointly measuring the commuting quadratures $R^Y_1, \ldots, R^Y_n$ defined in \eqref{eq:ry}. Then,
\begin{equation}\label{eq:enX-appendix}
\mathbb{E}_Y\left(\left|Y\right|^2 + \frac{1}{4}\left|\nabla\ln p(Y)\right|^2\right) \le \sum_{i=1}^{2m} \mathrm{Tr}\left[R_i^X\,\rho_X\,R_i^X\right] < \infty\,,
\end{equation}
where $p$ is the probability density of $Y$.
\end{lem}
\begin{proof}
By convexity of the Fisher information and linearity of the other terms, it is sufficient to prove \eqref{eq:enX-appendix} when $\rho_X$ is a pure state, \emph{i.e.}, $\rho_X = |\psi\rangle\langle\psi|$ for some unit vector $|\psi\rangle\in L^2(\mathbb{R}^m)$.

We can complete $B$ to a symplectic matrix, such that there exists a Gaussian symplectic unitary operator $U$ acting on $L^2(\mathbb{R}^m)$ such that $R_i^Y = U^\dag\,R_{2i-1}^X\,U$ for any $i=1,\,\ldots,\,n$ (see \emph{e.g.} [\onlinecite[Section 5.1.2]{serafini2017quantum}]). Without loss of generality, we assume therefore that $R_i^Y$ corresponds to the position operator $(R_i^Y \psi)(x) = x_i \,\psi(x)$. We also assume that $n = m$, the general case following since the left-hand side of \eqref{eq:enX-appendix} decreases when restricted to $n$ marginals, using in particular that the classical Fisher information $\rho_Y \mapsto \mathbb{E}_Y\,\frac{1}{4}\left|\nabla\ln p(Y)\right|^2$ is  convex and lower semicontinuous.

$Y$ has density
\begin{equation}
p(\mathrm{d}^my) = \left|\psi_i(y)\right|^2\mathrm{d}^my\,,  \qquad y \in \mathbb{R}^{m}\,,
\end{equation}
and the finite energy assumption reads
\begin{equation}
\int_{\mathbb{R}^m}\left(\left|x\right|^2 \left|\psi(x)\right|^2 + \left|\nabla \psi(x)\right|^2\right)\mathrm{d}^mx  = \sum_{i=1}^{2m} \left\|R_i^X|\psi\rangle\right\|^2 < \infty\,.
\end{equation}
The claim follows from the identity
\begin{equation}
\mathbb{E}_Y \left|Y\right|^2   = \int_{\mathbb{R}^m} \left|y\right|^2 \left|\psi(y)\right|^2 \mathrm{d}^my\, ,
\end{equation}
and the inequality
\begin{equation}
\mathbb{E}_Y \,\frac{1}{4}\left|\nabla\ln p(y)\right|^2 = \int_{\mathbb{R}^m} \left| \nabla \sqrt{p}(y)\right|^2 \mathrm{d}^m y  \le \int_{\mathbb{R}^m} \left| \nabla \psi(y)\right|^2 \mathrm{d}^m y\, ,
\end{equation}
having used that $\left| \nabla \sqrt{p}(y)\right| = \left|\nabla |\psi(y)| \right| \le |\nabla \psi(y)|$.
\end{proof}

\begin{lem}\label{lem:finite-entropy-classical}
Let $X$ be a random variable with values in $\mathbb{R}^n$ with density $p$ such that \eqref{eq:enX} holds. Then, $X$ has finite Shannon differential entropy.
\end{lem}
\begin{proof}
Let $\gamma \in \mathbb{R}^{n \times n}_{>0}$ be the covariance matrix of $X$. Since Gaussian random variables maximize the differential entropy for a fixed covariance matrix, it follows that
\begin{equation}
S(X) \le \frac 1 2 \ln  \det (\mathrm{e}\,\gamma) < \infty\, .
\end{equation}
Gross' logarithmic Sobolev inequality, see \emph{e.g.} inequality (4)  in [\onlinecite{carlen1991superadditivity}] with $t=2\pi$, entails that
\begin{equation}
\mathbb{E}_X \ln p(X) \le \frac 1 2 \, \mathbb{E}_X \left| \nabla \ln p(X)\right|^2 - n\,,
\end{equation}
so that $S(X) > -\infty$.
\end{proof}

\section{Entropy of Gaussian states}\label{app:SG}

\begin{lem}\label{lem:SGas}
Let $\alpha\in\mathbb{R}^{n\times n}_{>0}$ and $\beta\in\mathbb{R}^{n\times n}_{\ge0}$.
Then, for $t\to\infty$,
\begin{align}\label{eq:SG1}
S_G(\beta + t\,\alpha) &= \frac{1}{2}\ln\det\left(\mathrm{e}\,t\,\alpha\right) + O\left(\frac{1}{t}\right)\,,\\
\label{eq:SQ1}
S_Q(\beta + t\,\alpha) &= \frac{1}{2}\ln\det\left(\mathrm{e}\,t\,\alpha\right) + O\left(\frac{1}{t}\right)\,,\\
\label{eq:SG2}
S_G(\alpha + t\,\beta) &= \frac{\mathrm{rank}\,\beta}{2}\ln t + O(1)\,,\\
\label{eq:SQ2}
S_Q(\alpha + t\,\beta) &= \frac{\mathrm{rank}\,\beta}{2}\ln t + O(1)\,.
\end{align}
\eqref{eq:SQ1} and \eqref{eq:SQ2} require $n$ to be even, and \eqref{eq:SQ2} requires also $\nu_{\min}(\alpha + t\,\beta)\ge1/2$ for sufficiently large $t$.
\end{lem}
\begin{proof}
\begin{description}
\item[\eqref{eq:SG1}]
We have
\begin{align}
S_G(\beta+t\,\alpha) &= \frac{1}{2}\ln\det\left(\mathrm{e}\left(\beta + t\,\alpha\right)\right) = \frac{1}{2}\ln\det\left(\mathrm{e}\,t\,\alpha\right) + \frac{1}{2}\ln\det\left(I_n + \frac{\beta\,\alpha^{-1}}{t}\right)\nonumber\\
&= \frac{1}{2}\ln\det\left(\mathrm{e}\,t\,\alpha\right) + O\left(\frac{1}{t}\right)\,.
\end{align}
\item[\eqref{eq:SG2}]
We have
\begin{equation}
S_G(\alpha + t\,\beta) = \frac{1}{2}\ln\det\left(\mathrm{e}\left(\alpha + t\,\beta\right)\right) = \frac{1}{2}\ln\det\left(I_n + t\,\beta\,\alpha^{-1}\right) + O(1)\,.
\end{equation}
Let $r=\mathrm{rank}\,\beta$, and let $\lambda_1\ge\ldots\ge\lambda_r>0$ be the nonzero eigenvalues of $\beta\,\alpha^{-1}$ with multiplicity.
We have
\begin{align}
\frac{1}{2}\ln\det\left(I_n + t\,\beta\,\alpha^{-1}\right) &= \frac{1}{2}\sum_{i=1}^r\ln\left(1 + t\,\lambda_i\right) = \frac{r}{2}\ln t + \frac{1}{2}\sum_{i=1}^r\ln\left(\lambda_i + \frac{1}{t}\right)\nonumber\\
&= \frac{r}{2}\ln t + O(1)\,.
\end{align}
The claim follows.
\item[\eqref{eq:SQ1}]
We have $\nu_{\min}(\alpha)>0$, therefore for any $t\ge1/2\nu_{\min}(\alpha)$ we have
\begin{equation}
\nu_{\min}(\beta + t\,\alpha) \ge \nu_{\min}(t\,\alpha) = t\,\nu_{\min}(\alpha) \ge \frac{1}{2}\,.
\end{equation}
\autoref{lem:SG} implies
\begin{equation}
S_Q(\beta + t\,\alpha) = S_G(\beta + t\,\alpha) + O\left(\frac{1}{{\nu_{\min}(\beta + t\,\alpha)}^2}\right) = S_G(\beta + t\,\alpha) + O\left(\frac{1}{t^2}\right)\,,
\end{equation}
and the claim follows from \eqref{eq:SG1}.
\item[\eqref{eq:SQ2}]
We have from \autoref{lem:SG}
\begin{equation}
S_Q(\alpha + t\,\beta) = S_G(\alpha + t\,\beta) + O(1)\,,
\end{equation}
and the claim follows from \eqref{eq:SG2}.
\end{description}
\end{proof}

\begin{lem}\label{lem:SG}
We have for any $\gamma\in\mathbb{R}^{2m\times2m}_{>0}$ with $\nu_{\min}(\gamma)\ge1/2$
\begin{equation}
S_G(\gamma) - m\ln\frac{\mathrm{e}}{2} \le S_G(\gamma) - \frac{m}{4\,{\nu_{\min}(\gamma)}^2}\ln\frac{\mathrm{e}}{2} \le S_Q(\gamma) \le S_G(\gamma)\,.
\end{equation}
\end{lem}
\begin{proof}
Let $\nu_1,\,\ldots,\,\nu_m$ be the symplectic eigenvalues of $\gamma$.
We have [\onlinecite[Chapter 12]{holevo2019quantum}]
\begin{equation}
S_Q(\gamma) = \sum_{i=1}^m g(\nu_i)\,,
\end{equation}
where for any $\nu\ge1/2$
\begin{equation}
g(\nu) = \left(\nu+\frac{1}{2}\right)\ln\left(\nu+\frac{1}{2}\right) - \left(\nu-\frac{1}{2}\right)\ln\left(\nu-\frac{1}{2}\right)\,.
\end{equation}
We have for any $\nu\ge1/2$
\begin{equation}
\ln\left(2\,\nu\right) \le \ln\left(\mathrm{e}\,\nu\right) - \frac{1}{4\,\nu^2}\ln\frac{\mathrm{e}}{2} \le g(\nu) \le \ln\left(\mathrm{e}\,\nu\right)\,,
\end{equation}
therefore
\begin{equation}
\ln\prod_{i=1}^m2\,\nu_i \le \ln\prod_{i=1}^m\mathrm{e}\,\nu_i - \frac{m}{4\,{\nu_{\min}(\gamma)}^2}\ln\frac{\mathrm{e}}{2} \le S_Q(\gamma) \le \ln\prod_{i=1}^m\mathrm{e}\,\nu_i\,,
\end{equation}
and the claim follows since
\begin{equation}
\prod_{i=1}^m\nu_i = \sqrt{\det\gamma}\,.
\end{equation}
\end{proof}

\bibliography{aipsamp}

\end{document}